\documentclass[final,1p,times]{elsarticle}
\usepackage{amssymb}
\usepackage{amsmath}
\usepackage{amsthm}
\usepackage{mathrsfs}
\usepackage{amsfonts}
\usepackage{graphicx}
\usepackage{bm}
\usepackage[dvipsnames]{xcolor}
\usepackage{epsfig}
\usepackage{amsfonts}
\usepackage{bbold}
\usepackage[titletoc]{appendix}
\usepackage{multirow} % para las tablas
\usepackage{footnote}
\usepackage{blkarray}% http://ctan.org/pkg/blkarray
\makesavenoteenv{tabular}
\makesavenoteenv{table}
\usepackage{hyperref}
\usepackage[normalem]{ulem}
\biboptions{sort&compress}

\newcommand{\beq}{\begin{equation}}  
\newcommand{\eeq}{\end{equation}}  
\newcommand{\beqa}{\begin{eqnarray}}  
\newcommand{\eeqa}{\end{eqnarray}}

\newcommand{\nico}[1] {\color{black}}
\newcommand{\fer}[1] {\color{black}}
\newcommand{\redmark}[1] {\color{black}}
\newcommand{\bluemark}[1] {\color{black}}

\newcommand{\expSTMESR}{ \cite{Baumann_Paul_science_2015,
Natterer_Yang_nature_2017,Choi_Paul_natnano_2017,Willke_Paul_sciadv_2018,Yang_Bae_prl_2017,
Willke_Bae_science_2018,Bae_Yang_scadv_2018,
Willke_Singha_nanolett_2019,Willke_Yang_natphys_2019,Yang_Paul_prl_2019,Yang_Paul_science_2019,
Seifert_Kovarik_eabc_2020,Weerdenburg_Steinbrecher_arxiv_2020,Steinbrecher_Weerdenburg_arxiv_2020,
Veldman_Farinacci_arXiv_2021}}

\newcommand{\expESRFe}{\cite{Baumann_Paul_science_2015,Natterer_Yang_nature_2017,Choi_Paul_natnano_2017,Willke_Paul_sciadv_2018,
Willke_Bae_science_2018,Willke_Singha_nanolett_2019,Seifert_Kovarik_eabc_2020,Weerdenburg_Steinbrecher_arxiv_2020}}

\newcommand{\expESRTi}{\cite{Yang_Bae_prl_2017,
Willke_Bae_science_2018,Bae_Yang_scadv_2018,
Willke_Yang_natphys_2019,Yang_Paul_prl_2019,Yang_Paul_science_2019,
Seifert_Kovarik_pr_2020,Seifert_Kovarik_eabc_2020,
Weerdenburg_Steinbrecher_arxiv_2020,Steinbrecher_Weerdenburg_arxiv_2020,Veldman_Farinacci_arXiv_2021}}

\begin{document}
\begin{frontmatter}

\title{A theoretical review on the single-impurity electron spin resonance on surfaces}

\author[iudea,ull]{Fernando Delgado\corref{cor1}}
\ead{fernando.delgado@ull.edu.es}
\address[iudea]{ Instituto Universitario de Estudios Avanzados en Física Atómica, Molecular y Fotónica (IUDEA), Universidad de La Laguna }
\address[ull]{ Departamento de F\'{i}sica, Universidad de La Laguna, C/Astrof\'{i}sico Francisco S\'anchez, s/n. 38203, Tenerife, Spain}

\author[cfm,dipc]{Nicol{\'a}s Lorente}
\address[cfm]{Centro de F{\'{\i}}sica de Materiales
              CFM/MPC (CSIC-UPV/EHU),  20018 Donostia-San Sebasti\'an, Spain}
\address[dipc]{Donostia International Physics Center (DIPC),  20018 Donostia-San Sebasti\'an, Spain}

\begin{abstract}
The development of electron spin resonance (ESR) combined with scanning tunneling spectroscopy (STM)
is undoubtedly one of the main experimental breakthroughs in surface science of the last decade 
thanks to joining the extraordinarily high energy resolution of ESR (nano-eV scale) with the 
single-atom spatial resolution of STM (sub-\AA ngstr\"om scale).
While the experimental results have \fer{significantly} grown with the number of groups that have succeeded in implementing the technique, the physical mechanism behind it is still unclear, with several different mechanisms proposed to explain it. Here, we start by revising the main characteristics of the experimental setups and observed features. 
Then, we review the main theoretical proposals, with both their strengths and weaknesses. 
One of our conclusions is that many of the proposed mechanisms share the same basic principles,
\fer{the time-dependent electric field at the STM junction is modulating the coupling of the spin-polarized transport electrons with the local spin.} This explains why these mechanims are
 essentially equivalent in a broad picture. We analyze the subtle differences between some of them and how they compare with the different experimental observations. 
\end{abstract}


\end{frontmatter}
\date{today}

\tableofcontents

\section{Introduction}

Electron spin resonance (ESR, also known as electron paramagnetic resonance -EPR) and nuclear magnetic resonance (NMR) are two of the most successful and broadly used characterization and imaging techniques, employed from medical and bio applications to chemical characterization of paramagnetic complexes~\citep{Abragam_Bleaney_book_1970}. Among their strengths, they provide a high-energy resolution of magnetic transitions even in ambient conditions, limited basically by the time the wave functions {preserve their phase}, known as the {\em phase decoherence time} $T_2$, and non-invasive and non-destructive measurement, ideal for in-vivo applications. The spatial resolution of these techniques is essentially limited by the magnetic field gradients that can be implemented. Presently this implies commercial equipments~\cite{Fratila_Velders_arac_2011} capable of spatial resolutions of the order of 100 $\mu$m$^3$. While these resolutions can be enough for biomedical imaging applications, a higher spatial resolution is desirable for surface and chemical analysis, such as the one needed in quality assessment of thin-layer industrial applications (hard drives, magnetic tunnel junctions for sensing applications, etc). NMR with nm$^3$ spatial resolution has been demonstrated by {optical means} using a 
nitrogen-vacancy center in a nanodiamond at the tip apex of atomic force microscope (AFM)~\cite{Mamin_Kim_science_2013} {or, more recently, NV-center layer at the surface of a diamond chip~\cite{Glenn_Bucher_nature_2018}.

\begin{figure}[t]
\includegraphics[width=0.8\linewidth]{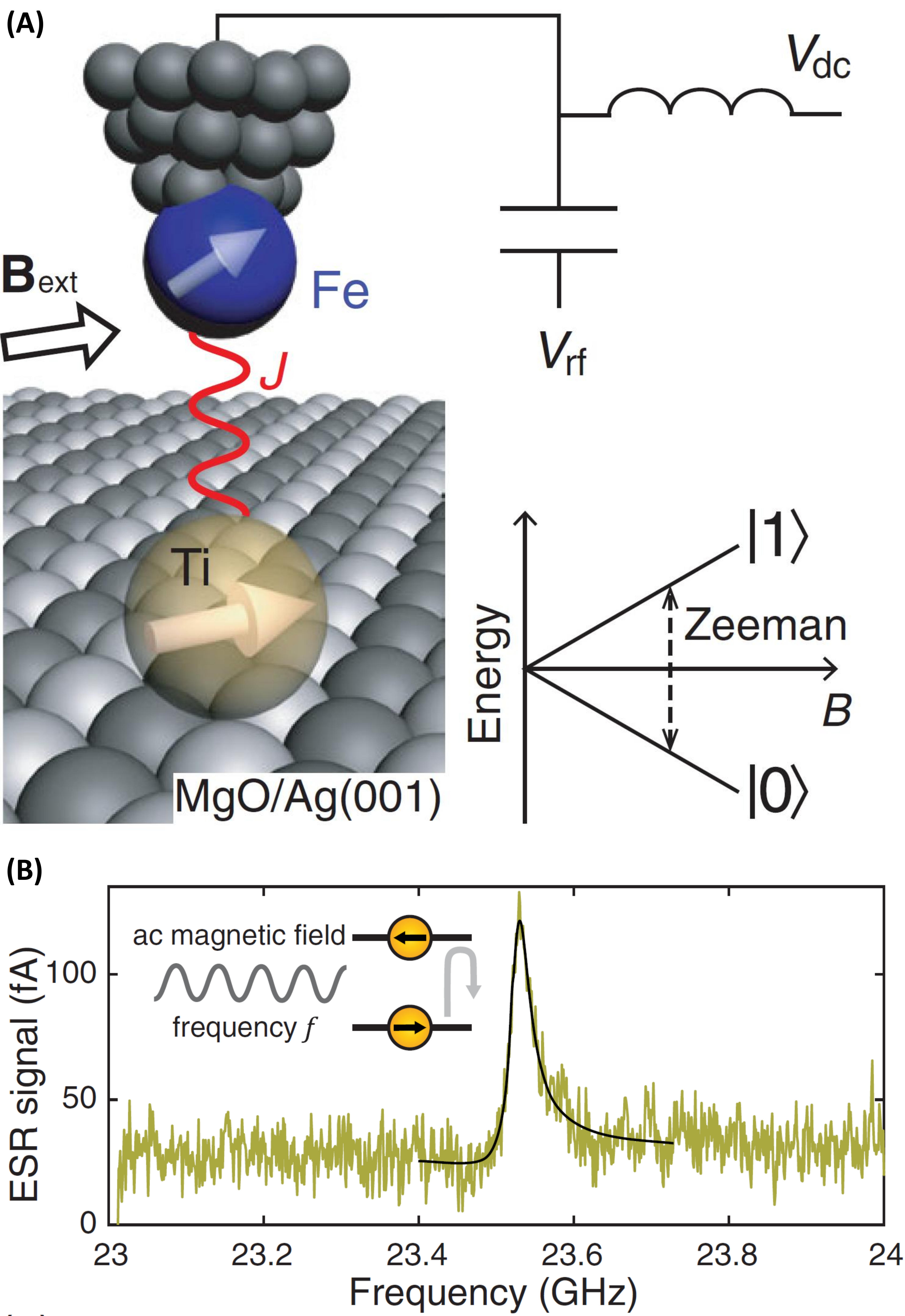}
\caption{(A) Experimental setup of the STM-ESR exemplified for the Ti  adatom on the MgO/Ag(001) surface. The time-dependent electric field at the STM junction gives place to a  (time-dependent)
coupling between the spin-polarized tip (in the displayed example, ended in an Fe atom),  and the local spin of the adatom. When the frequency of the $V_{RF}$ voltage maches the transition frequency, a characteristic resonant feature appears in the DC tunnel current. (B) Example of ESR spectrum ($\Delta I$) recorded on a Ti atom using continuous wave ESR (set point: $V_{DC}=$ 50 mV, $I_{DC}= 10$ pA, $V_{RF}= 20$ mV, $T= 1.2$ K).
Reprinted with permission from Yang {\em et al.}, Phys. Rev. Letts. {\bf 122}, 227203 (2019). Copyright (2019) by the American Physical Society.
\href{https://doi.org/10.1103/PhysRevLett.122.227203}{DOI:10.1103/PhysRevLett.122.227203}.
}
\label{scheme}
\end{figure}

Early attempts of ESR implemented on an STM (STM-ESR) used a radiofrequency rotating magnetic field
\fer{working at room temperature} to modulate the static magnetic field. 
When the central frequency of the narrow detection band equaled the
frequency of the signal, the detector showed a signal until the magnetic-field modulation frequency was out of the detection band.
These initial attempts demonstrated the presence of resonant magnetic field-dependent features, revealed  as an increased noise in the tunnel current at the spin precession 
frequency~\cite{Manassen_Hamers_prl_1989,Balatsky_Nishijima_advphys_2012}.
Unfortunately, difficulties \fer{in controlling} some of the experimental variables made the results sometimes irreproducible and unclear. 

All-electrical ESR with atomic spatial resolution has been a long-standing goal of the surface science community. 
The use of a radiofrequency-driven electric field renders the technique more simple to achieve.
In 2014, M\"ullegger {\em et al.} reported STM-ESR and NMR~\cite{Mullegge_Tebi_prl_2014} of a terbium-double decker molecule (TbPc$_2$) residing
inside the STM tunnel junction formed by the tip and an Au(111) substrate, and they argued that the observed  features of the resonance
did not satisfy well-established dipole selection rules ($\Delta J_z=\pm 1$) 
for the magnetic resonance~\cite{Mullegge_Rauls_prb_2015}. 
Baumann {\em et al.}~\cite{Baumann_Paul_science_2015} provided the first conclusive and fully-reproducible 
results of STM-ESR of single atoms. Among the many experimental challenges that Baumann {\em et al.} had to solve to achieve this reproducibility, the full characterization and compensation of the frequency-dependent transmission function of the STM junction was a key technical point. Since then, a plethora of experimental results of STM-ESR has appeared\expSTMESR, demonstrating the STM-ESR with a resolution capable of revealing hyperfine interactions on single transition-metal atoms deposited on MgO~\cite{Yang_Willke_natnano_2018}.
This technique has shown remarkable achievements such as the unambiguous measurement of the
magnetic moment of a single $f$-electron atom~\cite{Natterer_Yang_nature_2017}, the differentiation of individual Ti atoms by their number of nucleons~\cite{Willke_Bae_science_2018},
{or the full characterization of the spin interaction between two adatoms~\cite{Choi_Paul_natnano_2017,
Yang_Paul_prl_2019} among others. }

These all-electric STM-ESR works use a high-end
radiofrequency generator to modulate the applied bias between
tip and sample, see Fig.~\ref{scheme}. Another successful approach uses a second macroscopic
electrode positioned at milli-meters from the STM tip, working as a radiofrequency
antenna~\cite{Seifert_Kovarik_pr_2020,Seifert_Kovarik_eabc_2020}. Within the available data, these different approaches seem capable of the same
achievements.

The above results show the spectroscopic capabilities
of continuous-wave STM-ESR. Indeed, the long-time average  under a
continuous sinusoidal driving of the tip-sample electric field
is an efficient way of revealing Rabi oscillations between
local spin states. Beyond spectroscopy obtained in the
continuous-wave mode, it is interesting
to do STM-ESR with microwave pulses. Pulses of radiation
interacting with single spins are at the core of quantum operations
\cite{Mete_2006,Press_2008,Bayliss_2020}.  Recently, pulses of
radiofrequencies were mastered in the STM-ESR~\cite{Yang_Paul_science_2019},
which paves the way for coherent manipulation of single spins with
atomic resolution, see Fig.~\ref{fig2}.  

% Theory
Understanding the actual driving mechanisms of local spins in
the STM junction is not easy.  For a comprehensive review on the
theoretical approaches trying to explain the early
experiments by Mannassen {\em et al.}, see the work of Balatsky {\em
et al.}~\cite{Balatsky_Nishijima_advphys_2012}. 
In essence, the possible
mechanism lies into two non-mutually exclusive classes: those works
that rely on the spin-orbit coupling to explain the interaction between
the local spin and the tunnel current, or those where such a
coupling is not demanded, such as the direct exchange mechanism based
on a tunnel barrier modulation.
 Here, we will instead
focus on the theory of all-electrical STM-ESR
\fer{and the related theories that emerged after the seminal work by 
Baumann {\em et al.}~\cite{Baumann_Paul_science_2015}.}
%

% Trivial
 Direct estimations of the radiofrequency magnetic field created by the
alternating tunnel current~\cite{Yang_Paul_prl_2019} or the 
radiofrequency antenna~\cite{Seifert_Kovarik_pr_2020}
 displacement current permit us to discard this straightforward
origin. Moreover, recent
measurements of the STM-ESR dependence on the standoff tip-sample distance $d_0$
are not compatible with the dependence of these radiofrequency 
current-mediated mechanisms~\cite{Seifert_Kovarik_eabc_2020}.

\begin{figure}[t]
\includegraphics[width=1.\linewidth]{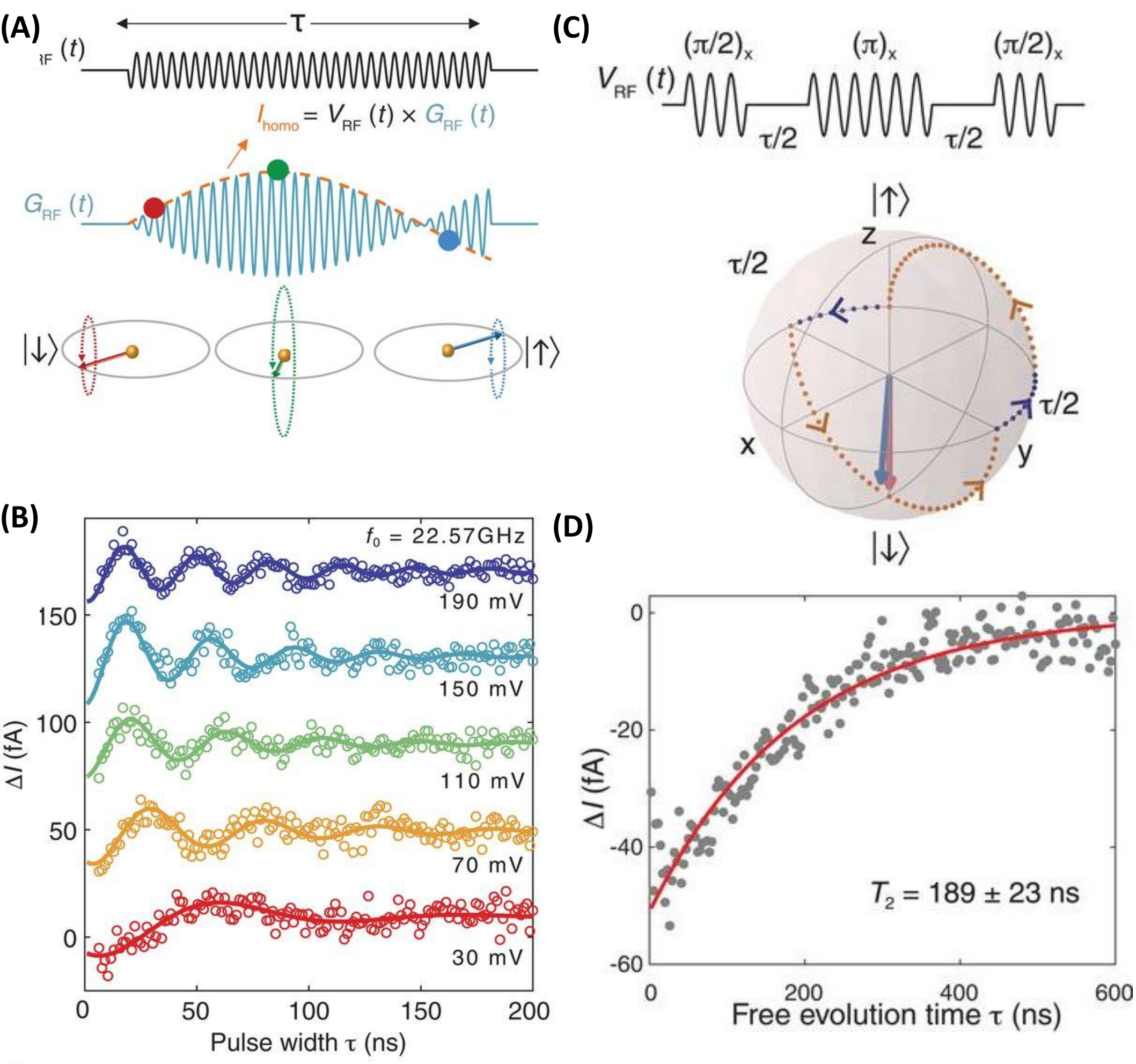}
\caption{Pulsed ESR protocols for coherent control of the spin. (A) Scheme depicting the STM-ESR detection mechanism for a pulsed sequence of radiofrequencies. The measured current $I_{\rm homo}$ results of the time-average product of $G_{RF}$, the instantaneous tunnel conductance when $V_{RF}$ is on, and $V_{RF}(t)$.
The pulse length and shape is chosen to force a coherent rotation of the spin on the Bloch sphere.
(B) Measured (dots) and fitted (solid lines) Rabi oscillations of a Ti spin at different values
of $V {RF}$, at the indicated frequencies ($V_{DC} = 60$ mV, $I_{DC} = 4$ pA, $B=
0.90$ T).  The data are offset
vertically for better visibility. 
(C) Pulse scheme (top) and Bloch sphere
representation (bottom) of the spin evolution in the rotating frame for Hahn-echo measurement.
(D) Measured (dots) and exponential fit (red line) of the spin-echo signals versus the free evolution time $\tau$ at $B=0.82$ T and frequency 20.55 GHz.
From Yang {\em et al.}, Science {\bf 366}, 509 (2019), 
\href{https://science.sciencemag.org/content/366/6464/509}{DOI:10.1126/science.aay6779}.
 Reprinted with permission from AAAS. 
}
\label{fig2}
\end{figure}

A different mechanism based on a radiofrequency current-induced magnetic field was proposed~\cite{Berggren_Fransson_2016}, 
where the spin-polarized
current was driving finite uniaxial and transverse anisotropy fields on
the local spin moment. However, this mechanism predicts null ESR signal
on half-integer spins, such as
the spin-1/2 STM-ESR on Ti that was observed later~\expESRTi.

%%%%% Piezoelectric mechanism
The piezoelectric displacement of the adatom produced by the radiofrequency
electric field in the STM junction is at the origin of several proposed
mechanism. Baumann {\em et al.}~\cite{Baumann_Paul_science_2015} argue
that this piezoelectric vibration of the adatom produces a modulation
of the single-ion magnetic anisotropy, which is seen in the spin-polarized DC tunnel
current thanks to its magnetoresistive dependence.
Another mechanism associated with the piezoelectric displacement is 
the modulation of the exchange and dipolar magnetic fields created by
the tip~\cite{Lado_Ferron_prb_2017,Seifert_Kovarik_eabc_2020}. Additionally, for
certain systems, the modulation of the $g$-factor due to the
variation of the crystal field felt by the local moment may be sufficient to
explain the ESR signal~\cite{Ferron_Rodriguez_prb_2019}.
These mechanisms will be discussed in Sec.~\ref{piezoR}. Furthermore,
there is additional mechanism that does not need to be
related to the piezoelectric coupling, but can be strongly enhanced
by the piezoelectric response of the substrate.
%%%% Cot.
This mechanism is based on the modulation of the tunneling
probability by the applied alternating bias voltage. This modulation can stand on either a piezoelectric deformation of the tunnel
barrier, and thus, a variation of the effective barrier width, or a change in the barrier height associated with the alternating
bias \fer{voltage} established at the STM junction. In both cases, its effect on the DC
tunnel current can be taken into account by a time-dependent description
of the cotunneling processes~\cite{Galvez_Wolf_prb_2019}.
%%%%% Others
To end this list, we mention a different proposal based on the
time-dependent spin-torque produced by the spin-polarized tunnel
current~\cite{Shakirov_Rubtsov_prb_2019}. 

All-electric STM-ESR has been very successful on local magnetic
moments adsorbed on a bilayer of MgO grown on Ag (100). 
To the best of our knowledge, 
this is the only substrate that {has led}  
to ESR-active
atomic and molecular adsorbates.  The strong piezoelectric response
of MgO layers is a good indication of the importance of mechanical
vibrations induced by the oscillating electric field of the tip in
the ESR mechanism. However, as emphasized by the cotunneling approach,
the ultimate origin may be the way the tunneling properties change with
the applied electric field. This modulation
 can be greatly enhanced by the piezoelectric response of the substrate, but it is in principle
independent of it. This fact implies that all-electric STM-ESR 
should be possible on a number of substrates besides MgO.

%%% Relation with optical ultrafast STM spectroscopy %%%%%%%%%%%%%%%%%
\fer{Although in this work we have concentrated on the all-electrical STM-ESR, it is worth pointing out that many of the theoretical approaches presented here can be easily applied to other experimental setups. 
Generally speaking, experimental techniques that provide nanometer spatial resolution with information on the fast sub-ns dynamics interrogate the properties of the individual atoms/molecules strongly coupled to an electromagnetic field by probing the light-matter interaction with techniques such as absorption~\cite{Moerner_Kador_prl_1989,Steeves_Elezzabi_apl}, Raman scattering~\cite{Yampolsky_Fishman_natpho_2014},
fluorescence induced by linear or nonlinear absorption~\cite{Orrit_Bernard_prl_1990,Plakhotnik_Walser_science_1996},
photothermal effects~\cite{Gaiduk_Yorulmaz_science_2010}, or photoelectron spectroscopy~\cite{Reutzel_Li_natcomm_2020}. 
In particular, time resolutions of the order of ps have been achieved by combinig femtosecond laser pulses with a scanning probe in photoconductively-gated STM~\cite{Yarotski_Averitt_joptsa_2002,Terada_Yoshida_natpho_2010}. %which requires specialized probed or sample structure and presents significant problems associated to the capacitive coupling effects~\cite{Groeneveld_Kempen_apl_1996}. 
An alternative successful approach to achieve sub-ps time resolution while keeping the nanometer spatial resolution has been demonstrated using ultrafast terahertz scanning tunneling microscopy achieved by coupling free-space terahertz pulses, which are generated by an ultrafast laser source, to an STM tip~\cite{Cocker_Jelic_natphot_2013,Cocker_Peller_nature_2016}. In these experiments, the terahertz pulses act like fast voltage transients that sample the $I(V)$ curve of the tunnel junction. Despite the significant differences with STM-ESR, in both cases we have that the local quantum system is under the action of an essentially monochromatic fast-oscillating driving field. Hence, the description of the time evolution of the driven system in Sec.~\ref{SecRabi} can be equally applied. This should not be surprising since the methodology used derives mainly from quantum optics. Transient-time information in STM-ESR would be accessible only in pulsed STM-ESR~\cite{Yang_Paul_science_2019} or combining STM-ESR with pump-probe techniques~\cite{Veldman_Farinacci_arXiv_2021}.
}

The structure of this article is a first section reviewing
qualitative aspects of STM-ESR together with the figures of merit
governing the spin-resonance process. Next, we review the different proposals
for the mechanisms behind STM-ESR. Followed by a section analyzing common
and specific features. 
The article ends with
a summary and conclusions section. Several appendices give more details
on the derivations of the involved theories.

\section{Basics of STM-ESR. Figures of merit.}

Electron spin resonance is a technique based on exciting
a Rabi oscillation between two magnetic levels in an electronic system.
It is then important to understand Rabi oscillations between two levels. 
In this section, we give a brief overview of the driven two-level system (TLS) and the figures
of merit that characterizes the stationary dynamics, typically
obtained as long-time averages. As we mentioned in the Introduction,
the specificities of STM-ESR is to use an all-electric
technique to drive a single atomic or molecular spin system
in an STM junction. Besides driving the spin,
we need to detect it. The ESR signal {translates into} 
a change
in electronic current when {either} the frequency of the driving bias voltage~\cite{Baumann_Paul_science_2015} or
of the radiofrequency field created by a microantenna~\cite{Seifert_Kovarik_eabc_2020}
matches the Larmor frequency of the TLS.
In this section, 
we briefly review these concepts and how they apply to the STM-ESR case.

\subsection{Driven spin system: the Rabi frequency \label{SecRabi}}

A key ingredient in ESR is the Rabi frequency, also known as Rabi flop rate. Just by
understanding and evaluating this quantity, much insight
can be gained in the resonant excitation of a spin. We briefly
review its definition and meaning in the context of the dynamics
of the driven TLS.

\subsubsection{The TLS}
The magnetic resonance of a quantum system can be described exactly in
the same way as standard ESR or nuclear
magnetic resonance (NMR), {and it is formally equivalent to an optically driven system~\cite{Cohen_Grynberg_book_1998}}. To begin with, let us assume that a quantum
system described by a Hamiltonian ${\cal H}_S$ is subjected to a time-dependent field oscillating at an angular frequency $\omega$.  Due to this time-dependent interaction,
we assume that only two states $|a\rangle$ and $|b\rangle$
with eigenenergies $\epsilon_a$ and $\epsilon_b$ of the unperturbed ${\cal H}_S$  are
connected by a transition frequency\footnote{In the context of magnetic resonance of a spin $S=1/2$ system, this will be the Larmor frequency $\omega_L=\gamma B$, with $\gamma$ the electronic gyromagnetic factor} $\omega_{0}\equiv \omega_{ba}=(\epsilon_b-\epsilon_a)/\hbar>0$ close
enough to the driving frequency $\omega$.  Under these conditions, the
driven isolated system can be treated as a TLS. Thus,
we will concentrate on the time evolution on the subspace of dimension two spanned by
$|a\rangle$ and $|b\rangle$.  If the radiative shifts of levels $a$
and $b$ are included in the 
frequency $\omega_{0}$, we can
write the driven system Hamiltonian as
\begin{eqnarray}
{\cal H}_S(t)&=&-\frac{\hbar\omega_{0}}{2}\hat \sigma_z+\hbar\Omega\cos(\omega t)\hat\sigma_x \nonumber \\
&=&
\begin{pmatrix}
-\frac{\hbar\omega_{0}}{2} & \hbar \Omega\cos (\omega t) \\
\hbar \Omega\cos (\omega t) & \frac{\hbar\omega_{0}}{2}
\end{pmatrix},
\label{CohTE}
\end{eqnarray}
where $\hat\sigma_\alpha$ is the $\alpha$-Pauli matrix in the
$\left\{ |a\rangle,|b\rangle\right\}$ basis set and $\Omega$ is
called the Rabi frequency, which characterizes the strength of the
driving.

To get some further insight, let us consider the stationary case with $\Omega=0$.  If the system is initially prepared in state $|\psi_0\rangle$, the state at time $t$ will be given by
\beq
|\Psi(t)\rangle=\langle a|\psi_0\rangle e^{-i\omega_0 t/2}|a\rangle+\langle b|\psi_0\rangle e^{i\omega_0 t/2}|b\rangle.
\label{psi_t}
\eeq 
In other words, the state of the system will oscillate coherently between $|a\rangle$ and $|b\rangle$ at a frequency $\omega_0/2$ (the probabilities or state populations oscillate at $\omega_0$). 

The dynamics of the driven system (\ref{CohTE}) is certainly more complex. In disentangling the role of the Rabi frequency, it is convenient to write the driving term as 
\beq
\frac{\Omega}{2}\left( {\cal J}_+ e^{-i\omega t}
+{\cal J}_-e^{i\omega t}+{\cal J}_- e^{-i\omega t} +{\cal J}_+ e^{i\omega t}
\right),
\label{RWA_terms}
\eeq 
where we have defined ${\cal J}_+ =|b\rangle\langle a|$ and ${\cal J}_-=|a\rangle\langle b|$. Thus,
the driving term contains resonant processes where the system is excited by the absorption of a photon (or decay by the emission of a photon), the first two terms in Eq. (\ref{RWA_terms}),  and non-resonant processes where it is excited after emitting a photon (or it decays after absorbing a photon), the last two terms in Eq. (\ref{RWA_terms})~\cite{Cohen_Grynberg_book_1998}. These last two non-resonant processes are very unlikely and can be then neglected. This constitutes the rotating-wave approximation (RWA). Hence, the driving term under the RWA takes the form $\hbar\Omega/2\left({\cal J}_+ e^{-i\omega t}+{\cal J}_- e^{i\omega t}\right)$.

The main benefit of the RWA is that it permits describing the system dynamics in a rotating frame where the time-dependence $e^{\pm i\omega t}$ can be removed.  
%The time evolution of the resulting driving terms can be easily removed. 
In particular, by applying the transformation to the rotation frame defined by $U=e^{-i\omega t/2\hat\sigma_z}$, we get that the Hamiltonian is transformed according to ${\cal H}_S(t)\to U{\cal H}_S(t)U^\dag+i \dot U U^\dag$, so that
\beq
{\cal H}_S^{\rm RWA}\approx -\frac{\hbar\delta \omega}{2}\hat\sigma_z+\frac{\hbar\Omega}{2}\hat \sigma_x,
\eeq
where $\delta \omega=\omega-\omega_0$. This Hamiltonian can be easily diagonalized to obtain the two eigenvalues, $\pm \tilde \Omega/2$,  where $\tilde\Omega= \sqrt{\Omega^2+\delta\omega^2}$ is known as the generalized Rabi frequency. This result allows us to interpret the Rabi frequency in quite simple terms. When the system is driven by a quasi-resonant perturbation of intensity $\hbar\Omega$, its state coherently oscillates at $\tilde \Omega/2$ in a frame rotating around the $z$-axis of Hamiltonian (\ref{CohTE}) at frequency $\omega$, very much like in Eq. (\ref{psi_t}). In particular, at resonance ($\delta \omega=0$), it will oscillate at half the Rabi frequency $\Omega/2$, i.e.,
the state $|\psi(t)\rangle$  will oscillate between states $|	a\rangle$ and $|b\rangle$  at a frequency determined by the \textit{strength} of the driving.

The proposed mechanism for STM-ESR must
be such that the Rabi frequency matches the experimental one. A
system will be ESR-active if we have sizeable Rabi frequencies.

\subsubsection{Bloch equations for the TLS}

The time evolution of the isolated system will be determined by the Schr\"odinger equation $i\hbar\partial_t \psi(t)={\cal H}_S(t)\psi(t)$, or equivalently, by the coherent evolution of its density matrix,  $i\hbar\dot { \rho}(t)=\left[{\cal H}_{S}(t),{ \rho}(t) \right]$. 
Let us now consider a {\em very large} system composed by our quantum system, characterized by a Hamiltonian ${\cal H}_S(t)$, and the rest, which we will call {\em reservoirs}. In particular, we will be interested in the limit where the reservoirs are very large systems with quasi-continuum spectra, in other words, thermal baths that will remain in thermal equilibrium. If $|\Psi(t)\rangle$ is the complete many-body time-dependent state
of the full system, the density matrix of the whole system will be ${\bm\rho}_T(t)= |\Psi(t)\rangle \langle \Psi (t)|$. Since in general we are interested only in the evolution of our quantum system, we can disregard most of the information of the baths. Hence, it is particularly useful to trace out the degrees of freedom of the reservoirs by introducing the {\em reduced density matrix} of the system:
\beq
\rho(t)={\rm Tr}_B\left[\rho_T(t) \right],
\eeq
where ${\rm Tr}_B[\dots]$ indicates the trace over the bath degrees of freedom, ${\rm Tr}_B[\dots]=\sum_B \langle B | \dots |B \rangle $.
The reduced density matrix can be used to describe the evolution of the system coupled to its environment,
 but the price to pay is that the Heisenberg equation of motion of $\rho(t)$ will contain, in addition to the coherent part $\left[{\cal H}_{S}(t),{ \rho}(t) \right]$, a very complex dissipative part accounting for the coupling with the baths. This is done
by including a super-operator called the Liouvillian, see for example
Ref.~(\cite{Breuer_Petruccione_book_2002}). In general, such a problem cannot be solved, and further approximations are needed. Fortunately, in many cases, the coupling of the quantum system with its environment can  be considered weak, so a second-order perturbative treatment on this coupling is sufficient. In addition, the bath correlation time is often very short, much shorter than any other characteristic time scale of the system, which allows us to neglect memory effects. This is known as the Markovian approximation. Under these conditions, one can derive a Bloch-Redfield master equation for $\rho(t)$ where the Liouvillian takes the form of the time-independent Lindblad super-operators~\cite{Breuer_Petruccione_book_2002}.

In order to
derive the main equations controlling the ESR, we will use the standard
approximation employed to derive the optical Bloch equations for the two
level system~\cite{Cohen_Grynberg_book_1998}. As such, we will assume
a relaxation rate $1/T_1$ and a decoherence rate $1/T_2$ independent
of the driving field. This approximation is valid if the effect of this
radiation can be neglected during the correlation time  $\tau_c$ of the
vacuum fluctuations responsible for the spontaneous emission, 
which holds
if $\Omega\ll \omega_0$\cite{Cohen_Grynberg_book_1998}.  In addition, we
limit the discussion to the TLS under the RWA where fast oscillating terms, at frequencies $2\omega_0$, are neglected. 

%The TLS can be mapped into a spin-1/2 system taking
$a$ and $b$ as spins down and up respectively. Then,
we can identify~\cite{Yosida}
the quantities ${\cal S}_+=\rho_{ba}e^{-i\omega
t}$, ${\cal S}_-=\rho_{ab}e^{i\omega t}$ and ${\cal
S}_z=(\rho_{bb}-\rho_{aa})/2$, and splitting the real and imaginary
parts by introducing ${\cal S}_x=({\cal S}_+ + {\cal S}_-)/2$ and ${\cal
S}_y=({\cal S}_+ - {\cal S}_-)/(2i)$, we get the following dynamical
equations\cite{Cohen_Grynberg_book_1998}
\beqa
\dot {\cal S}_x &=& \delta\omega {\cal S}_y -\frac{ {\cal S}_x}{T_2},
\crcr
\dot {\cal S}_y &=& -\delta\omega {\cal S}_x +\Omega {\cal S}_z -\frac{1}{T_2}{\cal S}_y,
\crcr
\dot {\cal S}_z&=& -\Omega {\cal S}_y -\frac{1}{T_1}\left({\cal S}_z +{\cal S}_z^{(0)}\right),
\label{OptBEq}
\eeqa
where ${\cal S}_z^{(0)}$ is the thermal equilibrium value of ${\cal S}_z$.
Equations (\ref{OptBEq}) are equivalent to the macroscopic dynamical equations for the magnetization  dynamics in ESR or NMR described in a rotating frame~\cite{Abragam_Bleaney_book_1970}, see Fig.~\ref{fig3}. In particular, the steady state solution of these effective $S=1/2$ spin components are given by
\beqa
{\cal S}_x &=&\frac{1}{2}\Omega \delta\omega T_2^2  /( 1+\xi^2+(T_2\delta\omega)^2)
\crcr
{\cal S}_y &=&\Omega T_2 %{\cal F}%(\delta\omega,\Omega,T_1,T_2),
/( 1+\xi^2+(T_2\delta\omega)^2)
\crcr
\frac{({\cal S}_z -{\cal S}_z^{(0)})}{{\cal S}_z^{(0)} }&=&-\Omega^2 T_1T_2 %{\cal F}
/( 1+\xi^2+(T_2\delta\omega)^2),
\label{steadyBeq}
\eeqa
where  $\xi^2=T_1T_2\Omega^2$.

\begin{figure}[ht]
\includegraphics[width=0.9\linewidth]{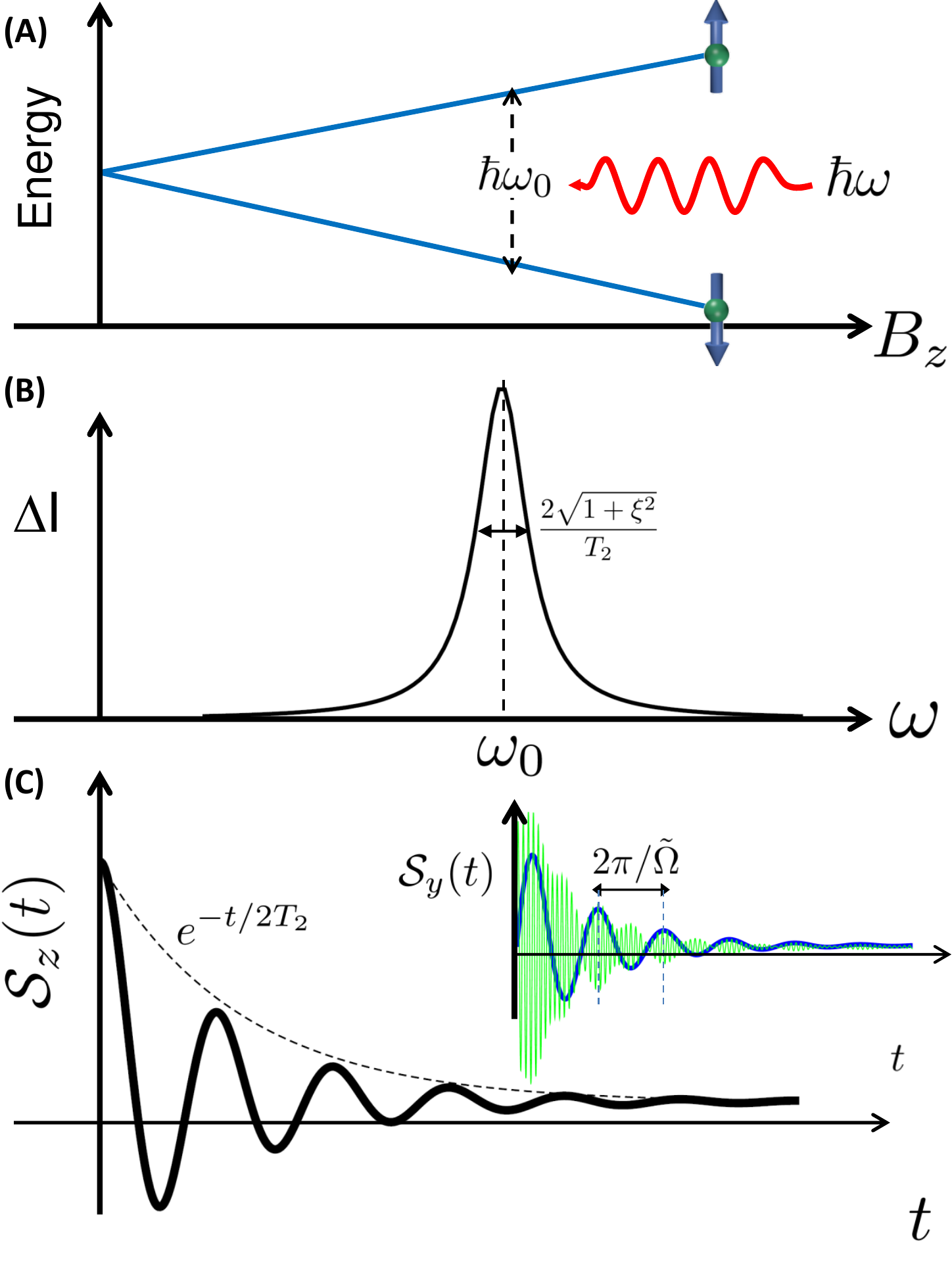}
\caption{ESR of a two level system. (A) Energy level scheme of a TLS split by a magnetic field $B_z$ with transition frequency $\omega_0$. (B) When a radiation of frequency $\omega$ close to $\omega_0$ impinges the quantum system, it can induce a resonant signal. In the STM-ESR, the DC tunnel current displays a resonant feature centered on $\omega_0$. (C) According the to Bloch equations, the longitudinal effective magnetization, directly related to the occupation difference, %Eq. (\ref{OptBEq}), 
decays exponentially with a characteristic time $1/(2T_2)$, experiencing a precession matching
the generalized Rabi frequency, $\tilde\Omega$. For the
present set of parameters ($1/T_1\lesssim 1/T_2\lesssim  \Omega\ll \omega_0 $), the decay of the longitudinal
magnetization ${\cal
S}_z=(\rho_{bb}-\rho_{aa})/2$
 is not controlled by $T_1$, but
by the decoherence time, $T_2$, see Eq. (6) of Ref. (\cite{Gauyacq}). The transverse components ${\cal S}_{x/y}$, defined by the coherence $\rho_{ab}(t)$,  also experience a damped oscillation, see inset. If observed on a rotating frame (thick blue line), the oscillation  has frequency $\tilde\Omega$ but, when observed in a lab frame (thin green line), it oscillates with the much higher frequency $\omega_0$.
}
\label{fig3}
\end{figure}

The dynamics of the TLS in the steady-state or at long times are then completely controlled by three quantities proper to the TLS and
its environment: the Rabi frequency, $\Omega$, the relaxation time, $T_1$, and the decoherence time, $T_2$.
The meaning of the two relaxation times, $T_1$ and $T_2$, can be easily derived from (\ref{OptBEq}).
 $1/T_1$ is de exponential decline of the population {difference}.
 It is indeed a relaxation time {towards its equilibrium value}.
$1/T_2$ controls the non-diagonal terms \fer{$\rho_{ab}(t)$} of the density matrix (coherences)\fer{, 
which} account for the
 relative phase between the states of ${\cal H}_S$. \fer{Notice that, in terms of our ficticious ${\cal S}=1/2$ spin, $T_2$ corresponds to the decay time of the transversal spin components ${\cal S}_x,{\cal S}_y$, see Eq. (\ref{OptBEq}).
 } This phase is destroyed if an event introduces
a random phase in one of the wavefunction components, like scattering with an external particle. This is a purely dephasing event, and it is usually assigned a pure-dephasing rate $1/T_2^*$. 
 But of course, destroying the population of one of the state also leads to
losing the coherence between states. Then the decoherence rate {of a TLS} is given by
\beq
\frac{1}{T_2}=\frac{1}{2T_1}+\frac{1}{T_2^*}.
\eeq
Depending on the actual parameters of the system,
the dynamics of the populations can actually vary. For the \fer{example depicted in} Fig. \ref{fig3}, we find that the difference
in populations or longitudinal magnetization, $S_z$, actually decays as $1/2T_2$. This type of dependence
of the population difference on the decoherence rate has been explored in Ref. (\cite{Gauyacq}). \fer{By contrast, for the commonly found case in magnetic adatoms where $T_2\ll T_1$~\cite{Delgado_Rossier_prl_2012,Delgado_Rossier_pss_2017}, the evolution of the coherences is much faster than the occupation one, so one can consider the coherences at each time corresponding to a {\em steady-state condition} corresponding to the value of the population difference. }

In summary, these three numbers, $\Omega$, $T_1$ and $T_2$, control the features of  continuous-wave ESR where the driving is a single harmonic function of time.
Typical figures for experiments on a Fe adatom on MgO/Ag(100)~\cite{Baumann_Paul_science_2015,Willke_Paul_sciadv_2018} are
$\Omega \sim 10^{6}-10^7$ s$^{-1}$, $T_1\sim 10^{-6}-10^{-4}$ s, $T_2\sim 10^{-7}-10^{-8}$ s. We see that decoherence {dominates} the evolution.
Indeed, $\sim60\%$ of the tunneling electrons cause decoherence~\cite{Willke_Paul_sciadv_2018} of the spin of Fe.

\fer{Here, we have focussed on the interpretation of the Bloch equations for a TLS as a fictitious $S=1/2$ spin and its connection with the time evolution of the magnetization in macroscopic ESR. Nevertheless, the validity of this treatment goes beyond the TLS, and it also applies to other research fields, being quantum optics the most eminent~\cite{Cohen_Grynberg_book_1998}. In particular, they are commonly used to describe the interaction of an atom with monochromatic radiation, typically a laser field. Since the laser-based detection scheme allows accessing the very short time scales, they can be used to interpret the transient time-evolution. Moreover, they permit analyzing the radiative forces exerted by the light beam on either static or moving atoms. }

\subsection{Detection mechanism of STM-ESR \label{Secdetec}}

In STM-ESR, the magnetic resonance is measured through the tunneling magnetoresistance effect at the STM junction. This means that the junction conductance depends on the relative alignment between the tip magnetic moment, \fer{which defines the $z$-direction,} and the precessing adatom spin, i.e., a contribution proportional to $\vec m^T\cdot  \langle\vec  S\rangle=  m^T_z\langle S_z\rangle$, where $\vec m^T$ is the magnetic moment of the tip. Hence, it is not surprising that the change in the tunnel current $\Delta I$ between the RF on and RF off follows the frequency dependence of ${\cal S}_z$ in Eq. (\ref{steadyBeq})~\cite{Baumann_Paul_science_2015}.

The DC current %measured at the STM junction 
can be written in terms of the tunnel rates $\Gamma^{TS(ST)}_{nn'}$  defined as the transition rate between the $n$ and $n'$ states of the magnetic atom ($a$ and $b$ if we stick to the TLS approximation) when an electron from the tip (surface) is scattered to the surface (tip). If we write the voltage and frequency-dependent occupation of the $n$ state as $P_n(V,\omega)$, the tunnel current can be written as
\beq
I(V,\omega)=e\sum_{nn'}P_n(V,\omega)\left( \Gamma_{nn'}^{TS}-\Gamma_{nn'}^{ST}\right).
\eeq
The occupations $P_n(V,\omega)$ will differ in general from the thermal equilibrium ones, and they will be given by the stationary solution of the dynamical equation of the density matrix, i.e., $\dot{\bm \rho}(t)=0$.  Using the solution of the Bloch equations for the TLS, Eq. (\ref{steadyBeq}), one can write the occupation difference between the two states connected by the ESR~\cite{Delgado_Rossier_pss_2017,Galvez_Wolf_prb_2019}
\beq
P_b(V,\omega)-P_a(V,\omega)=\delta P_0\left(1+\frac{\xi^2}{1+\xi^2+\delta\omega^2 T_2^2}\right),
\label{Occpdiff}
\eeq
where $\xi^2=\Omega^2T_1T_2$ and $\delta P_0$ will be given by the non-equilibrium occupation difference in the absence of RF modulation. In the case of a TLS, a closed expression of the tunnel current can be obtained with $\delta P_0$ {given by} %\nico{approximated by} %corresponding to 
the thermal equilibrium occupation difference between the $b$ and $a$ states. 

Let us consider the experimentally relevant case where a large DC bias voltage is applied ($|eV|\gg k_BT,\hbar\omega_0$). Under these conditions, $\Gamma_{a,b}\approx \Gamma_{b,a}$. 
In addition, we will assume a dominant elastic contribution to the current, $\Gamma_{a,a}^{TS}\approx \Gamma_{bb}^{TS}\gg\Gamma^{TS}_{a,b}$, which is quite frequently found in spin-polarized STM experiments~\cite{Delgado_Rossier_prb_2010}. Then, by introducing the expectation value of the spin state, $S^z_{aa}=-S^ z_{bb}\equiv -{\cal S}$ and the quantity $\zeta=T/J$, where $T$ is the direct spin-independent tunneling coupling and $J$ the tunneling exchange coupling (also known as the potential and exchange scattering matrix elements respectively), the change in tunneling current is
\beq
%I_{\rm On}(V,\omega)-I_{\rm Off}(V) %\approx G_j\left(1+\delta P(V,\omega){\cal S}\zeta {\cal P}_T\right)V,
\Delta I
\approx
G_j\delta P_0 {\cal S}\frac{\xi^2\zeta{\cal P}_T}{1+\xi^2+\delta \omega^2 T_2^2},
\label{Ifreq}
\eeq
where ${\cal P}_T$ is the tip polarization and $G_j$ the junction elastic conductance.
{Notice that since $|b\rangle$ is the excited state, $\delta P_0\le 0$, so $\Delta I\le 0$ ($\Delta I\ge 0$) when the tip and the adatom magnetic moment are aligned (antialigned). In other words, we can expect a deep (peak) in the DC tunnel current for the aligned (antialigned) tip polarization when the resonant condition is satisfied.}

Equation (\ref{Ifreq}) clearly illustrates the detection mechanism. The DC current is \fer{sensitive to both} the 
non-equilibrium induced adatom magnetic moment $\delta P_0{\cal S}$ and the tip moment \fer{(proportional to 
the spin polarization ${\cal P}_T$}), which explains by STM-ESR is not observed in the absence of spin-polarization.

Experimental data of STM-ESR under larger  RF
voltages~\cite{Yang_Paul_prl_2019,Seifert_Kovarik_eabc_2020,Seifert_Kovarik_pr_2020,Willke_Paul_sciadv_2018} 
and/or larger tunnel conductances~\cite{Willke_Singha_nanolett_2019}
 shows important deviation from the symmetric frequency profile derived from Eq. (\ref{Ifreq}). This asymmetric line shape has been explained in terms of the \textit{homodyne} contribution to the DC current~\cite{Yang_Bae_prl_2017,Bae_Yang_scadv_2018,Yang_Paul_science_2019} (homodyne refers to processes
happening at the same frequency). In the present context, there is a new contribution to the long-time average or DC current due to the RF component of the applied bias voltage. The symmetric contribution is due to the above magnetoresistive mechanism that yields a change in current with precessing frequency for a DC applied bias voltage. The homodyne contribution is, however, due to time-dependent tunneling conductance associated with the spin precession when it matches
the RF applied bias, and it is responsible \fer{for} the appearance of an asymmetric Fano-like profile with frequency. 
In essence, the measured sinusoidal lock-in signal contains an approximately linear background as a function of the driving frequency that depends on both the Rabi frequency and the junction conductance. In the case of continuous STM-ESR, the resulting amplitude of the ESR signal close to resonance can be written as~\cite{Bae_Yang_scadv_2018,Seifert_Kovarik_eabc_2020}
 \beq
  \Delta I=I_0+I_p(\xi)\left(\cos\alpha+\delta\omega\sin\alpha \frac{ T_2\,V_{RF}}{2\,\Omega \, T_1\,V_{DC}} \right),
 \label{DeltaFano}
 \eeq
where $I_0$ is a background contribution that accounts for the non-linearity of the junction, $\alpha$ is the angle between the tip magnetic moment and the rotating magnetic moment,
and $I_p(\xi)$ defines the amplitude of the ESR signal, given by
the equivalent of Eq. (\ref{Ifreq}) for the total current,
\beq
I_p(\xi)=I_{\rm sat}%\frac{\xi^2}{\sqrt{1+\xi^2}}(1+\beta\sin\varphi).
\frac{\xi^2}{1+\xi^2+\delta\omega^2 T_2^2}.
\eeq
Here we have introduced the saturation current $I_{\rm sat}$ as measured experimentally~\cite{Bae_Yang_scadv_2018}, while the phase shift between the RF excitation and the local spin has been neglected~\cite{Seifert_Kovarik_eabc_2020}.

Interestingly, Eq. (\ref{DeltaFano}) contains both a symmetric contribution associated to the tunneling magnetoresistance ($\sim \cos\alpha$) and the asymmetric homodyne contribution  ($\sim V_{RF} \sin\alpha$).  
Thus, the resonance profile can be characterized by three quantities~\cite{Seifert_Kovarik_eabc_2020}: the amplitude $I_p(\xi)$, the width
%\beq
$\Gamma=(\pi T_2)^{-1} \sqrt{1+\xi^2}$ (we recall that $\xi^2=\Omega^2T_1T_2$)
%\eeq
and the asymmetry of the profile controlled by $( T_2V_{RF})/(2\Omega T_1V_{DC})$.

Interestingly, one can explain the asymmetric line shape without resorting to the homodyne detection scheme. G. Shavit {\em et al.}~\cite{Shavit_Horovitz_prb_2019} demonstrated that the origin of the deviations from the symmetric line shape stands on the frequency dependences of the dissipative dynamics neglected in the derivation of the Bloch equations.
By using a series of unitary transformations of the system Hamiltonian and avoiding the usual secular approximation with regard to the low-frequency Rabi frequency, they derived  the asymmetric line shape (\ref{DeltaFano}). Moreover, their approach allowed them to account for the signal asymmetry sign depending on the applied bias, magnetic field orientation, and driving amplitude.
This more involved approach does not contradict the above results. It rather goes beyond the simplifications  above to express in a succinct formulation the same physics.

\section{Proposed driving mechanism }

\subsection{Piezoelectric response \label{piezoR}}
The piezoelectric displacement of the adatom driven
by the radio frequency (RF) tip-induced electric field is at the core of several mechanisms proposed to explain the STM-ESR. For instance, the original mechanism based on the modulation of the crystal field~\cite{Baumann_Paul_science_2015} relies on this piezoelectric effect. In particular, for the 4-fold symmetric Fe/MgO/Ag system, Baumann and coworkers argue that the Rabi oscillations were proportional to~\cite{Baumann_Paul_science_2015} 
\beq
\Omega\cos(\omega t)\propto  F(t)\langle a|\left(L_+^4+L_-^4\right)|b\rangle.
\eeq
Here, $F$ is a crystal field parameter mixing the orbitals states that differs in four units of orbital angular momenta. The basic principle behind this expression is the following. The radiofrequency
voltage induces a displacement $z(t)=z_0\cos(\omega t)$ of the surface-adatom distance, much smaller than the equilibrium tip-adatom distance $d$. Then, this change gives place to a RF modification of the crystal field parameter $F(t)$. Although the proposed mechanism correctly reproduces the magnetic field dependence of the Rabi flop rate for the Fe/MgO/Ag system~\expESRFe, it fails to explain the observed resonances
 on other adatoms, such as the Ti-H ($S=1/2$)~\expESRTi or the NMR of Cu~\cite{Yang_Willke_natnano_2018}.

In addition to the renormalization of the crystal field parameters due to the piezoelectric displacement, this oscillation modifies the exchange coupling between the adatom spin and the itinerant electrons~\cite{Lado_Ferron_prb_2017}. The coupling with both tip and surface electrons is then modified{, and thus, the tunneling-related exchange coupling.} 
In particular, the finite tip magnetic moment can give place to an effective magnetic field that, due to the RF piezoelectric oscillation, it follows the electric field. This mechanism is analyzed in detail in Sec.~\ref{JoaquinT}.

Moreover, there is a third mechanism associated to piezoelectric displacement. In addition to the renormalization  of the crystal field parameters, the piezoelectric displacement also gives place to a  renormalization of the $g$-factor. However, Ferr\'on {\em et al.} estimated that this contribution to the Rabi flop rate is between one and two orders of magnitude smaller than the experimental values.~\cite{Ferron_Rodriguez_prb_2019}

In general, the efficiency of this piezoelectric effect is determined by the amplitude of the induced oscillation and the parametric dependence on the displacement. By assuming a small displacement, we can write~\cite{Lado_Ferron_prb_2017}
\beq
{\cal H}_S(z)\approx{\cal H}_S^{(0)}+z (t) \left.\frac{d {\cal H}_S}{d z}\right|_0.
\label{Hpiezo}
\eeq
This expression evinces how an RF oscillation of the adatom-surface distance gives place to a driving term of the adatom Hamiltonian. In particular, if the driving frequency $\omega$ is close enough to the Larmor frequency $\omega_{0}$ between states $|a\rangle$ and $|b\rangle$ of ${\cal H}_S^{(0)}$, the induced Rabi flop rate  will be
\beq
\Omega_{\rm piez.}=\frac{z_0}{\hbar}\langle a|\left.\frac{d {\cal H}_S}{d z}\right|_0|b\rangle.
\eeq

\subsubsection{Exchange modulation \label{JoaquinT}} %:{ Fernandez-Rossier THEORY}
A universal mechanism based on a modulation of the exchange coupling between the tip and impurity magnetic moments for STM-ESR was proposed by J. Lado and coworkers~\cite{Lado_Ferron_prb_2017}. The basics idea is as follows. In an STM-ESR setup, a spin-polarized tip is used, which will give place to an exchange coupling between the localized spin of the adatom and the tip magnetic moment. If this exchange is modulated in time at the driving frequency by some given mechanism, it will give place to a time-dependent exchange, i.e., $J^T(t)=J_0^T+\delta J^T\cos(\omega t)$. 

The effect of this exchange is equivalent to an effective magnetic field, 
\beq
\vec B_{\rm eff}(t)=-\frac{J^T(t)}{(g\mu_B)^2}\vec m^T=\vec B_{\rm eff}^0+\delta \vec B_{\rm eff}\cos(\omega t),
\label{Befft}
\eeq
 where $\vec m^T= m^T \hat n$ is the tip-magnetic dipole moment. Notice that here we have assumed a static magnetic moment of the tip. Later we will come back to this assumption. Hence, the working principle of the proposed mechanism is evident. If the total DC magnetic field $\vec B_0= \vec B_{\rm ex}+\vec B_{\rm eff}^0 $ is applied along a given direction, which we will denote as $\parallel$, %any residual magnetic anisotropy,
and it gives place to a splitting $\hbar\omega_0$ of the energy levels $|a\rangle$ and $|b\rangle$,\footnote{The splitting $\hbar\omega_0$ can be induced by a combination of magnetic anisotropy terms defined along some crystallographic direction $z'$, and the total static field $\vec B_{\rm eff}^0 $. For instance, in the case of the Fe adatom on MgO/Ag, where
the tip magnetic moment is aligned with the large static field along the in-plane direction $x'\parallel \hat n$, the splitting $\Delta$ is induced by a small out-of-plane $z'$ field $B_z\ll B_x$, while the large $B_{x'}$ field only produces a minor energy shift due to the large out-of-plane magnetic anisotropy of the system.}
the time-dependent modulation of the exchange coupling translates into a Rabi flop rate $\Omega =g\mu_B  \delta \vec B_{\rm eff}\cdot \vec S_{a,b}/\hbar$. Notice that one can only induce a finite Rabi flop rate if the direction $\hat n$ of the RF field, which is parallel to the tip magnetic dipole moment $\vec m^T$,  is such that $\hat n\cdot \vec S_{a,b}/\hbar\ne 0$.
 For instance, in the case of the isotropic $S=1/2$ spin whose energy levels are split by a field along the $z$ direction, this is only possible if $\hat n\times \hat z\ne 0$. Hence, it works in the same way as the usual EPR and NMR experiments, where the applied static and RF magnetic fields are perpendiculars.
 
 Despite the intuitive meaning of the effective field, Eq. (\ref{Befft}), the source of the time-modulated exchange coupling
remains unclear. In the STM-ESR experiments, the only time-dependent field is the electric field between the tip and the sample generated by the radiofrequency bias voltage.\footnote{Notice that a modulation voltage in the KHz range may be applied to measure the $dI/dV$ through a locking technique, but this frequency is much lower than the relaxation and decoherence rates, so that it can be considered as a static field.} 
 J. Lado and coworkers~\cite{Lado_Ferron_prb_2017} proposed a piezoelectric origin for the modulation $\delta J^T$.   The radiofrequency electric field generates an oscillation of the tip-adatom distance, which due to the exponential dependence of the exchange  coupling, translates into an observable variation  of the 
spin-spin interaction that can efficiently drive the local spin. 
Then, they claimed that this modulation
changes the occupation of the surface spin states, 
which changes the average magnetic moment. Finally, the magnetoresistive response of the junction induced by the finite spin-polarization of the STM tip permits the observation of the resonant effect in the DC current.

Following Ref.~(\cite{Lado_Ferron_prb_2017}), let us assume that the applied  %$V_{RF}\cos(\omega t)$ 
voltage induces a displacement $z(t)=z_0\cos(\omega t)$ of the surface-adatom distance much smaller than the equilibrium tip-adatom distance $d$. The exchange coupling with the tip is assumed to decay exponentially, i.e., $J^T(z)=J_0^Te^{-(d_0-z)/\lambda}$, with $d_0=d+z$ the tip-adatom equilibrium distance. Then, by making a Taylor expansion we have that $\delta J^T\approx z_0 d\left.J^T(z)/dz\right|_{z=0}= J^T_0z_0/\lambda$.

As it was assumed in Eq. (\ref{Befft}), the quantum fluctuations of the magnetic
moment of the apex atom were ignored~\cite{Lado_Ferron_prb_2017}. These fluctuations are quenched by the combination of an applied magnetic field and strong Korringa damping~\cite{Korringa_physica_1950} with
the tip electron bath.  Therefore, the tip spin is treated in a 
mean-field or classical approximation, as done, for instance, in Ref.~(\cite{Yan_Choi_natnano_2015}).

The direct dipolar coupling between the tip magnetic moment and the adatom was also explored in Ref.~(\cite{Lado_Ferron_prb_2017}), finding that $\delta J^T_{\rm dip}\approx z_0 d \left.J^T_{\rm dip}(z)/dz\right|_{z=0}= 3z_0 J^T_{dip}/d $. Here, it was assumed that the tip magnetic moment was perpendicular to the tip-adatom axis.

In any of these proposed mechanisms, the Rabi frequency depends linearly on the adatom-surface oscillation amplitude, $z_0$. By taking the same sensible values of the tip-adatom distance ($d\approx  0.6$ nm), decay length ($\lambda\approx 0.06$ nm) and using the exchange coupling $J_0^T\approx 2$ meV estimated from experiments on Fe/Cu$_2$N/Cu(100)~\cite{Yan_Choi_natnano_2015}, it was estimated that $\delta J^T/z_0\approx 66.7$ meV/nm, while $\delta J^T_{\rm dip}/z_0\approx 0.02$ meV/nm. Hence, it was concluded that the Rabi frequency induced by the modulation of the dipolar coupling is negligible in the typical STM-ESR conditions. 

Density functional calculations (DFT) combined with electronic multiplet calculations 
allowed Lado \textit{et al.} to derive
effective magnetic anisotropy parameters for the Fe/MgO/Ag system, 
and the matrix elements $S^\alpha_{a,b}$ ($\alpha=x,y,z$) 
required in the calculation of the Rabi frequency~\cite{Lado_Ferron_prb_2017}. 
Using DFT, they estimated the amplitude of the adatom oscillation by assuming a harmonic restoring  potential with a spring constant $k$.
By doing so, they obtained a spring constant $k\approx 600$ eV/nm$^2$ for the Fe adatom. This translated into a displacement $z_0=0.044$ pm at $V_{RF}=8$ meV, or, equivalently, an exchange modulation $\delta J^T\approx 2.9\;\mu$eV. 
The predicted displacement was in good qualitative agreement with the displacement extracted by fitting the experimental Rabi frequency. 

However, Yang {\em et al.}  later measured the exchange coupling between a Ti-H molecule on MgO/Ag and the STM tip~\cite{Yang_Paul_prl_2019}.  They inferred a piezoelectric displacement of 2.9$\pm 0.2$ pm at $V_{RF}=10$ mV, which was a factor 40 larger than the one calculated using DFT. Thus, they argued that there should be other sources of piezoelectric displacement.
 Specifically, the relative vertical displacement
of the Ti atom with respect to the tip could be enhanced by
the motion of the MgO layer and/or the motion of the Fe atoms constituting the tip apex.

\subsubsection{Time-dependent local magnetic field \label{SecPietro}}

Equation (\ref{Befft}) clearly illustrates a simple idea for the driving mechanism: the spin system evolves under an effective magnetic field that has both a static and a time-dependent component. In the exchange modulation mechanism of Sec. \ref{JoaquinT}, the effective RF magnetic field was associated with the exchange coupling with the tip magnetic moment. However, this is not the only source of magnetic fields. For instance, the RF tunneling current and the RF displacement currents are expected to create an associated RF magnetic field. Yet, the corresponding magnetic fields have been estimated and discarded by their tiny contributions~\cite{Yang_Paul_prl_2019,Seifert_Kovarik_pr_2020}. In addition to the exchange field, the tip also induces a dipolar magnetic field whose intensity, direction and sign will depend on the relative orientation between the tip magnetic moment and the tip-adatom distance. This field was shown to be present for some STM tips~\cite{Willke_Yang_natphys_2019,Seifert_Kovarik_eabc_2020}. Seifert {\em et al.}~\cite{Seifert_Kovarik_eabc_2020} found that for a broad range of experimental conditions, their STM-ESR  signals could be accounted for by considering a (fixed-sign) exchange coupling and  a dipolar field of the form
\beq
 \vec B_{\rm eff}=\left( B_{\rm xc}^x+B_{\rm dip}^x\right)\hat {\bf x}+\left( B_{\rm xc}^z+B_{\rm dip}^z\right)\hat {\bf z},
\eeq
where $\hat{\bf x}$ and $\hat{\bf z}$ are unitary vectors along the directions x and z. Here, the exchange components could be described by an exponentially decaying amplitude, $B_{\rm xc}^{x/z}\propto B_0e^{-d/\lambda}$, while the dipolar field was given by $B_{\rm dip}^x=-b\sin\alpha/d^3$ and $B_{\rm dip}^z=2b\cos \alpha/d^3$. 
 Notice that the coordinate system has been chosen such that  $B_{\rm dip}^y=0$. They found that, in the case of Fe, for their tip realization $\alpha\approx 64^\circ $, $B_0^{\rm Fe}=-0.6\pm 0.1$ T (the sign was the same for all their tip realizations), $\lambda^{\rm Fe}=370\pm 60$ pm, and $ b^{Fe}=(0.2\pm 0.03)\mu_0\mu_B$. This indicates a clear competition between the dipolar and exchange fields, in contrast with the case reported before~\cite{Yang_Bae_prl_2017,Yang_Paul_prl_2019,Willke_Singha_nanolett_2019}. 
 
 Crucially, the tip-adatom distance dependence is different for both contributions, which will have important consequences for the Rabi frequency derived by the piezoelectric displacement, see Eq. (\ref{Hpiezo}). In particular, $\Omega_{\rm dip}\propto z/d_0^4$, while  $\Omega_{xc}\propto e^{-d_0/\lambda} z_0/\lambda$. 

Following previous works~\cite{Lado_Ferron_prb_2017,Yang_Paul_prl_2019}, Seifert {\em et al.}~\cite{Seifert_Kovarik_eabc_2020} computed the piezoelectric displacement $z$ from the structural response  of the adatom to an electric field using DFT. \fer{Since the adatom's local vibrational modes were found in the several-THz frequency window, they argue that the adatom will adiabatically follow the GHz electric field,} with displacements of the order of 0.5 pm/(V/nm) for both Fe and Ti-H. Compared to the $\sim 3.3$ pm/(V/nm) found by Lado {\em et al.}\cite{Lado_Ferron_prb_2017} or the 125 pm/(V/nm) found by Yang {\em et al.}\cite{Lado_Ferron_prb_2017,Yang_Paul_prl_2019} these values are much smaller, showing the difficulty in assigning clear physical causes to the STM-ESR origin.

A careful examination of the Rabi frequency dependence with the standoff (tip-adsorbate) distance $d_0$ showed a non-monotonous behavior for Fe and a monotonic decay for Ti-H~\cite{Seifert_Kovarik_eabc_2020}. Whereas the \fer{latter} can be explained considering only the exchange contribution, the highly anisotropic Fe case requires taking into account both tip-induced magnetic fields. Moreover, the presence of the specific magnetic anisotropy of Fe allows a finite driving Rabi frequency associated to the RF magnetic field along the direction of the static field, in clear contrast to the isotropic $S=1/2$ case of Ti-H and the usual ESR/NMR field configurations~\cite{Abragam_Bleaney_book_1970}.

\subsection{Radiofrequency excitation of spins via phonons}

Soon after the first reports of STM-ESR~\cite{Mullegge_Tebi_prl_2014},  it was clear that the  well-established dipole selection rules ($\Delta J_z=\pm 1$) did not describe the observed magnetic resonance~\cite{Mullegge_Rauls_prb_2015}. In 2015, M\"ullegger {\em et al.} proposed the following physical mechanism~\cite{Mullegge_Rauls_prb_2015}.  The RF electric field established at the STM junction was responsible for a periodic transient charging of the magnetic molecule, inducing an electric polarization that translates into an asymmetric deformation of the confinement potential. 
The resulting structural perturbation takes the form of a mechanical oscillation that permits angular momentum
transfer between the molecule's spin and its mechanical backbone, very much like in spin-phonon coupling. The substrate underneath lowers the symmetry, and it increases the 
molecule strain,  making the mechanism very
efficient.
Thus, in the proposal of M\"ullegger {\em et al.}~\cite{Mullegge_Rauls_prb_2015}, the internal mechanical degrees of freedom of the magnetic molecule  
where driving a spin-phonon-like coupling.

They argued that the efficient higher
spin excitations via one-step excitation processes in STM-ESR
pointed towards the involvement of mechanical degrees of
freedom as the way to fullfil the fundamental angular momentum
conservation.
Notice that as recently demonstrated for a single Fe adatom on Cu$_2$N atop N~\cite{Rejali_Coffey_npjqm_2020}, such an apparent violation of the total angular momentum can appear {in single adatoms} 
due to the coupling of the spin and orbital degrees of freedom. 

In particular, for  the magnetic anisotropy of the Terbium double-decker (TbPc$_2$) single-ion magnet on a Au(111) studied by  M\"ullegger {\em et al}~\cite{Mullegge_Tebi_prl_2014}, the resulting spin-phonon coupling takes the form
\beq
{\cal H}_{\rm s-p}=\sum_{\alpha\beta,\mu\nu}\Lambda_{\alpha\beta}^{\mu\nu} \frac{\partial u_\alpha}{\partial r_\beta}S^\mu S^\nu
\label{Hsp}
\eeq
where $e_{\alpha\beta}=\frac{\partial u_\alpha}{\partial r_\beta}$ ($\alpha\beta=x,y$) is the the transverse local strain, with the parameters $\Lambda_{\alpha\beta}^{\mu\nu}$ depending on the single ion anisotropy, phonon frequency and molecule moment of inertia~\cite{Chudnovsky_Garanin_prb_2005}. Equation (\ref{Hsp}) implies that the resulting spin-phonon coupling relaxes the dipolar selection rules, allowing transitions with $\Delta S_z=\pm 1$ and $\Delta S_z=\pm 2$.

The driving mechanism proposed for the TbPc$_2$ strongly relies on the mechanical degrees of freedom of the molecule. Hence, it is not clear whether a similar mechanism could be at work in the case of single magnetic adatoms. Even if the orbital degrees of freedom of the adatom were playing supplying the required angular momentum, it would not explain why the magnetic resonance is visible for bias voltages way below the orbital angular momentum excitation threshold~\cite{Rejali_Coffey_npjqm_2020}. Moreover, it seems that this mechanism works even in the absence of spin-polarization, in clear contrast to the experiments~\expSTMESR. Moreover, the standoff distance dependence observed experimentally~\cite{Seifert_Kovarik_eabc_2020} for Fe and Ti-H on MgO/Ag could not be explained by this deformation potential.

A different spin-phonon-mediated mechanism has been presented by Calero and Chudnovsky
\cite{calero_rabi_2007}.
Here, a propagating surface phonon shakes the adatom, inducing a time-dependent perturbation
of its magnetic environment.
Their calculations assume a single phonon displacing an adatom as a wave would displace a buoy in the sea. This changes the
normal of the atom with respect to the surface, inducing a torque on the adatom spin. This is different
from the above piezoelectric effect where the adatom-tip distance is modulated due to the action of the electrical field.
The Rabi frequencies thus computed only depend
on the phonon frequency, the atomic displacement of the surface atoms during the phonon propagation, the value of the oscillating spin as well as the transversal sound velocity.
When evaluated with reasonable values corresponding to Fe adsorbates on MgO, the experimental order-of-magnitude for Rabi frequencies is retrieved. However, the calculated Rabi frequencies scale as $\Omega\propto \omega^2$ where $\omega$ is
the driving angular frequency. This scaling would imply a strong change of signal as the frequency of the resonance changes that does not seem to be present in the available experimental data.

The above calculations\cite{calero_rabi_2007} were performed
assuming surface phonons propagating in one sense, which would correspond to a system that presents asymmetries along the surface. 
However, for the symmetry of adsorbed atoms, a piezoelectric displacement, as the one assumed before, is the only one that
retains the full symmetry of the problem. Indeed, this type of motion can be decomposed in phonon contributions, but as expected,
the effect of the transversal phonons cancel out, and the torque on the spin is identically zero, leading to zero Rabi frequencies.
In other words, in the presence of a surface system with cylindrical symmetry
about the magnetic adatom, the combined effect of all surface phonons perfectly cancels. This seems to be the case of most experiments, where the excitation of Rabi oscillations is
independent of the distance to defects or other symmetry-breaking features of the surface. Then, this mechanism does not seem plausible for realistic systems where a 
good degree of cylindrical symmetry exists about the magnetic adatom.

\subsection{Cotunneling theory\label{cotSec}}

The physical properties of diluted magnetic impurities in alloys have been studied in terms of both the Anderson model~\cite{Anderson_pr_1961} and the Kondo $s-d$ exchange model~\cite{Kondo_ptp_1964}. Both models have been extended and applied to magnetic adatoms on surfaces~\cite{Delgado_Rossier_pss_2017,Ternes_pss_2017,Choi_Lorente_rmp_2019}.
 In particular, when the adatoms are adsorbed on thin insulating surfaces, as it is the case in STM-ESR, the charge fluctuations included in the Anderson model and relevant in Hund's metals~\cite{Khajetoorians_Valentyuk_nn_2015} can be neglected, and both models are equivalent as soon discovered by Schrieffer and Wolff~\cite{Schrieffer_Wolff_pr_1966}. Thus, one would expect that the exchange mechanism for STM-ESR pointed out in Ref.~(\cite{Lado_Ferron_prb_2017}) could be rephrased in terms of the Anderson model. 
 Still, when mapping both models, the physical origin of the time-depending modulation of the exchange coupling, or equivalently, the effective time-dependent quantum magnetic field induced by the STM-ESR~\cite{Yang_Paul_science_2019}, can be completely different.

The idea behind the proposed cotunneling mechanism for STM-ESR is the following~\cite{Galvez_Wolf_prb_2019}. The RF electric field at the STM junction induces a change in the transmission amplitude $\tau(E)$ through the junction, a change that, to lowest order in the applied bias voltage, can be considered linear in $V_{RF}$ and proportional to the hopping amplitudes in the Anderson model. As it is well known in the context of open quantum systems, the coupling of a quantum system (QS) with a bath induces a renormalization of its energy spectrum. Thus, as a result of the time-oscillating hopping amplitudes there would be an effective oscillating term of the QS Hamiltonian, which, in principle, may contain finite off-diagonal matrix elements in the bases of eigenstates of the isolated QS. Henceforth, the Rabi frequency will be directly proportional to the off-diagonal matrix elements of this effective renormalization due to the RF oscillation of the transmission through
the tunnel barrier.

When the coupling between a magnetic impurity and the itinerant electrons is weak enough, one can treat this coupling effect using perturbation theory. 
For the magnetic adatoms on thin decoupling layers, second~\cite{Delgado_Rossier_pss_2017} or third ~\cite{Ternes_njp_2015} order perturbation theory is enough to reproduce almost all the observed experimental features, from the inelastic steps in the $dI/dV$ in IETS to the behavior of the relaxation and decoherence times. Perturbative treatment of an electronic multiplet model, a natural extension of the Anderson single orbital model, has been successfully applied to describe the spin-flips of magnetic atoms on Cu$_2$N~\cite{Delgado_Rossier_prb_2011}. More recently, this cotunneling theory was used to explain the complete reversal of the atomic unquenched orbital moment~\cite{Rejali_Coffey_npjqm_2020} of Fe adatoms on a Cu$_2$N/Cu$_3$Au(100) surface or the observed STM-ESR signal of different $3d$-transition metal atoms on MgO~\cite{Galvez_Wolf_prb_2019}.

Suppose a quantum system (QS), initially isolated and described by a Hamiltonian ${\cal H}_S$,  is weakly coupled to several metallic electrodes by a tunneling Hamiltonian
${\cal H}_t=\sum_{\alpha,{\bf i}} \left(V_{\alpha,{\bf i}} f^\dag_{\alpha}+h.c.\right)$, where ${\bf i}$ is a composite index ${\bf i}\equiv(\ell,\sigma)$ including all the quantum numbers in the QS, and $\alpha$ ($\alpha'$) stands for all the quantum numbers characterizing the free electrons, i..e., $\alpha\equiv (\eta,\vec k,\sigma$), with $\eta$ denoting the electrode, $\vec k$ the wave vector and $\sigma$ the spin direction, while $f_{\alpha}$ and $f^\dag_{\alpha}$ are the corresponding annihilation and creation operators.\footnote{We assume that the tunneling Hamiltonian conserves the spin, i.e., $\sigma_{\alpha}=\sigma_{{\bf i}}$.}
In the absence of time-dependent modulations, the time evolution of the coupled QS can be represented by an {\em effective cotunneling Hamiltonian} where the QS operators and the reservoirs operators have been factorized, and where the transition amplitudes can be easily evaluated by applying the Fermi Golden Rule to this effective coupling~\cite{Delgado_Rossier_prb_2011,Galvez_Wolf_prb_2019}. In other words, the total system formed by the QS coupled to the electrodes can be modeled by a Hamiltonian of the form ${\cal H}_S+{\cal H}_R+H_{\rm cot}$, where ${\cal H}_R$ is the Hamiltonian corresponding to the free-electron gases of the electrodes and $H_{\rm cot}$ is the effective cotunneling Hamiltonian that can be written as~\cite{Galvez_Wolf_prb_2019}
\beq
H_{cot.}\approx \sum_{\alpha\alpha'}
\left[\hat{T}_+(\alpha\alpha')f_\alpha^\dag f_{\alpha'}+\hat{T}_-(\alpha\alpha')f_\alpha f_{\alpha'}^\dag\right]
\label{Hcotun_eff}
\eeq
%\end{widetext}
 Here 
$\hat{T}_\pm(\alpha\alpha')$ denotes operators on the Hilbert space of ${\cal H}_S$ that can depend parametrically on $\alpha,\alpha'$. They can be related to transition amplitude matrix elements. 

The dynamics of such a coupled system can be described by the standard Bloch-Redfield  (BRF) master equation for weakly coupled QS fluctuating around a zero expectation value of the interaction. Thus, in general, we need to define a dressed system Hamiltonian ${\cal H}_S'={\cal H}_S+{\rm Tr}_R\left[H_{\rm cot.}\right]$ and interaction Hamiltonian 
 $H_{\rm cot.}'=H_{\rm cot.}-{\rm Tr}_R\left[H_{\rm cot.}\right]$, where ${\rm Tr}_R[\dots]$ denotes the trace over the reservoirs degrees of freedom. The resulting markovian BRF master equation takes the usual form of a Liouville's equation where the dissipative part reads as 
linear Lindblad
 super-operator ${\cal L}$ responsible for dissipation and thus,
 decoherence and relaxation. 
 
 Using the expressions (\ref{Hcotun_eff}) and some basic fermionic algebra, one can write down the effective coupled-system Hamiltonian as ${\cal H}_S'={\cal H}_S+\Delta {\cal H}_S$, where
\beq
\Delta {\cal H}_S= \sum_\alpha \hat{T}_+(\alpha\alpha)n_\alpha
+\sum_\alpha \hat{T}_-(\alpha\alpha)\left(1-n_\alpha)\right),
\eeq
where $n_\alpha=f_\alpha^\dag f_\alpha$ is occupation operator, which can take the values 0 or 1.

The above description of time-independent cotunneling cannot be directly applied when there is some time modulation in the total Hamiltonian, for instance, the modulation produced by an RF electric field in STM-ESR. As discussed in Ref.~(\cite{Galvez_Wolf_prb_2019}), the time-dependent electric field could, in principle, affect the three terms of the total Hamiltonian. While physical solutions are obtained only when the occupation of the single-particle states of the reservoirs are time-independent~\cite{jauho_wingreen_prb_1994}, both the hybridization couplings $V_{\alpha,{\bf i}}$ and the energy spectrum of ${\cal H}_S'$ could be affected by the time-dependent electric field. 
In the STM-ESR experiments \fer{performed} up to now, where the substrate consist of a MgO bilayer on top of Ag(100) surface, the electric field is almost fully screened by the MgO bilayer and, as demonstrated in Ref.~(\cite{Wolf_Reina_jpca_2020}), the energy spectrum of the magnetic adatoms are mostly unaffected by the external electric fields typically induced in the STM junction. Hence, this leads to $V_{\alpha,{\bf i}}(t)$ as the only (or dominant) source of time modulation. 

The hopping amplitudes $V_{\alpha,{\bf i}}$ of the Anderson model are directly related to the transmission amplitude $\tau(\epsilon)$ of the STM junction. Thus, it will depend on the different parameters that characterize the junction and, in particular, on the applied bias voltage. By making a series expansion around the Fermi energy of the electrodes, we can write %
%\beq
$V_{\alpha,{\bf i}}\approx V_{\alpha,{\bf i}}^{(0)}+V\left. \partial_V V_{\alpha,{\bf i}} \right|_{0}$.
%\eeq
%
By introducing the density of states $\rho(\epsilon)=\sum_{\alpha}\delta(\epsilon-\epsilon_\alpha)$, we can write the transmission amplitude as
 $\tau(\epsilon)=\sum_{\bf i}\tau_{\bf i}(\epsilon)\equiv {\cal C}\rho_\eta(\epsilon) V_{k_\epsilon\eta,{\bf i}}$ with ${\cal C}$ a normalization constant. Since the typical metal density of states can be considered constant in a energy window of a hundred meV or smaller around the Fermi level, we can take the value of $\rho$ at the Fermi surface. Thus, by making the corresponding substitution we have that 
\beq
V_{\alpha,{\bf i}}\approx V_{\alpha,{\bf i}}^{(0)}
\left(1+
%V\frac{1}{\tau_i(\epsilon)} \left. \frac{\partial \tau_{\bf i}(\epsilon) }{\partial V}\right|_{0}\right)
V\left. \frac{\partial \ln \tau_{\bf i}(\epsilon) }{\partial V}\right|_{0}\right).
\label{Vexpbarr}
\eeq
\begin{figure}[t]
\includegraphics[width=0.9\linewidth]{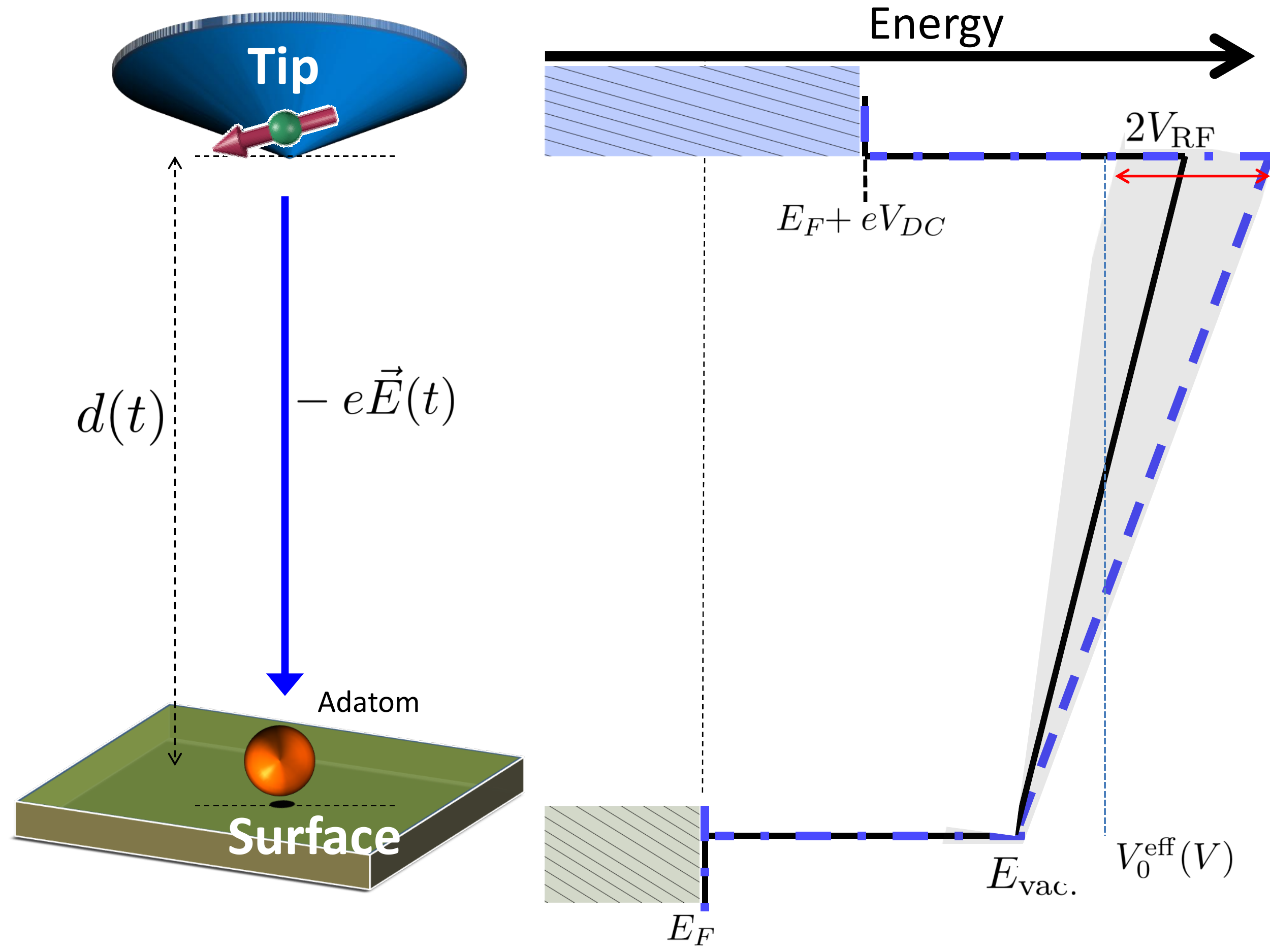}
\caption{Barrier modulation mechanism for the STM-ESR in the cotunneling description.
The time-dependent electric field $\vec E(t)$ at the STM junction
 give place to a variation of the elastic barrier transmission. In a simple description, the STM junction can be described by an effective square barrier whose high $V_0^{\rm eff}(V)$ depends on the applied bias voltage. The hopping matrix elements $V_{\alpha,{\bf i}}$ in the Anderson model are related to the barrier transmission, see Eq. (\ref{Vexpbarr}). Notice that the piezoelectric displacement mechanism also leads to a modulation of the barrier width, and hence, the transmission $\tau(\epsilon)$, with the consequent modulation of the hoppings $V_{\alpha{\bf i}}$.
}
\label{scheme}
\end{figure}

A quite accurate description of the transmission amplitude in an STM junction is given in terms of the Tersoff-Hamann model~\cite{Tersoff_Hamann_prb_1985}. In this approach,  the tip is modeled as a geometrical point, so that the tip effect can be simplified and factorized out of the problem, with the tunneling amplitude proportional to the surface wavefunction at the tip position and inversely proportional to the decay constant $\kappa$. 
An even simpler description of the STM junction can be provided by a one-dimensional square barrier at zero-bias voltage, assuming the same tip and surface work-functions, which evolves into a tilted barrier at finite bias voltage. Under this description, one can obtain a simple relation of the hybridization constants and the energy-dependent transmission amplitude. % $V_{\alpha}\propto \tau(E)$. 
Strictly speaking, under a finite bias, we should solve the problem of a triangular barrier, which involves the Airy functions~\cite{Fowler_Nordheim_prs_1928} and cumbersome expressions of the transmission amplitudes $\tau(E)$. \fer{A simpler qualitative description can be obtained using an effective square barrier at finite bias voltage where the applied bias voltage modifies the height of the barrier $V_0$.}
 This approach was used in Ref.~(\cite{Galvez_Wolf_prb_2019}) where it was shown that the transmission amplitude could be expressed as $\tau(\epsilon)\approx \tau_0(\epsilon)\left(1+\delta\tau(\epsilon) eV\right)$, where $\tau_0(\epsilon)$ is the transmission amplitude in the absence of bias voltage.
\beq
\delta\tau(E_F)\approx
\frac{-m^*L}{\hbar^2
\kappa_F}   .
\label{tausb}
\eeq
where we have considered an opaque barrier and taken the values at the Fermi level, i.e., $\kappa_F=\sqrt{2m^* (V_0-E_F)}/\hbar$, with  $m^*$ the effective mass and $L$ the tip-surface distance. This simple dependence is also reproduced in the Tersoff-Hamann model~\cite{Lounis_arXiv_2014}.
Here we should make the following remark. 
This result is based on the assumption that the effective barrier height $V_0$ changes linearly with the applied bias voltage $V$, i.e., $V_0(V)=V_0-\alpha V$. The coefficient $\alpha$ will depend on the electrostatic characteristics of the tip-sample junction. In the above results, we have assumed that the drop in the potential induced by the bias voltage occurs at the tip, while the surface's chemical potential is unaffected by the bias voltage. This asymmetric situation is common in STM junctions, while for quantum dots coupled to similar electrodes, the drop would occur symmetrically in the tip and surface chemical potentials, leading to $\alpha=0$.
Given the highly nontrivial potential dependence found by DFT in the MgO/Ag(100) surface under a finite electric field~\cite{Wolf_Reina_jpca_2020}, an accurate description of the effect of the electric field on the transmission may require to use a numerical estimation of the transmission function for a given self-consistent potential $V_{DFT}(\vec r,V)$.

 In any case, the central result is that we can relate the hybridization constants and the applied RF bias voltage to write down
\beq
V_{\alpha,{\bf i}}(t) \approx V_{\alpha,{\bf i}}^0\left(1+eV_{RF}\delta\tau(E_F)\cos(\omega t)\right).
\label{Vtd}
\eeq

A formal solution to the transport problem in the presence of a driving field requires using the Floquet theory in combination with the Keldysh Green’s function~\cite{Baran_Domanski_prb_2019}. However, a much simpler description can be used for weak coupling and weak driving regimes. Here the essential assumption is that the dissipative part of the time evolution of the density matrix, which is described by a Lindblad super-operator~\cite{Breuer_Petruccione_book_2002}, is not affected by the driving field. This approach is exactly the same used for the derivation of the optical Bloch equation of Sec.~\ref{SecRabi}. In that case, the system evolves under an effective time-dependent Hamiltonian
\beqa
{\cal H}_S'(t)&\approx&{\cal H}_S+\Delta {\cal H}_S^{(0)}+{\cal V}_{\rm cot}\cos(\omega t)
\label{Hcottd}
\eeqa
where we have defined
\beqa
{\cal V}_{\rm cot.}&=&
eV_{RF}\delta\tau(E_F)\sum_\alpha \left[ \hat{T}_+(\alpha\alpha)\hat n_\alpha
+\hat{T}_-(\alpha\alpha)\left(1-\hat n_\alpha)\right)\right].
\crcr &&
%\hbar \Omega_{\rm cot.}
\eeqa
A direct comparison of Eqs. (\ref{CohTE}) and (\ref{Hcottd}) allows us to identify the Rabi frequency extracted from the cotunneling model, as examined in detailed in Ref.~(\cite{Galvez_Wolf_prb_2019}).
First, we assumed that %\nico{Larmor} 
{transition} frequency $(E_b-E_a)/\hbar$ is the only one close to the driving frequency $\omega$.
Mathematically this can be expressed as follows. If $T_1(n)$ is the lifetime of state $n$, and $\omega_{nn'}=(E_n-E_{n'})/\hbar$, for any pair of states $(n,n')$ different from $a$ or $b$,  $(\omega-|\omega_{nn'}|)^2 T_1^2(n)|\gg 1$. Then, one directly obtains by direct inspection that $\hbar\Omega_{\rm cot}=\langle a|{\cal V}_{\rm cot}|b\rangle$, where we have defined the global phase of the interaction such that $\Omega \ge 0$.

\subsubsection{Relation with the {Kondo} exchange mechanism }
As mentioned at the beginning of Sec.~\ref{cotSec}, the single-orbital Anderson model can be mapped into the $S=1/2$ Kondo exchange coupling using the Schrieffer and Wolf canonical transformation. This procedure becomes far from trivial in the case of a multiorbital model. Thus, here we take an alternative mapping where our main goal will be to extract the relation between the  exchange coupling $J_{\eta\eta'}$ of a single impurity and the virtual transition operators $\hat T_\pm(\alpha\alpha')$. For this purpose, we omit the potential scattering term
and write the Kondo coupling as $V_K^{\rm ex}=\sum_{\eta\eta'}  J_{\eta\eta'}\sum_{a=x,y,z} S^{a} s^a(0)$, where $\vec s(0)$ is the spin density at the impurity site and $J_{\eta\eta'}$ corresponds to the exchange of an electron scattering from the $\eta$ to the $\eta'$  electrodes. Thus, introducing the Pauli matrices ${\bf \tau}^a$, we can write
\beqa
V_K^{\rm ex}
&=&\sum_{\eta\eta'}  \frac{J_{\eta\eta'}}{4}\sum_{\sigma}
\Bigg[\left[
\tau^+_{\sigma\bar\sigma}S^- +\tau^-_{\sigma\bar\sigma}S^+ 
\right]
\sum_{\vec k\vec k'}f^ \dag_{\vec k\eta\sigma}f_{\vec k'\eta'\bar\sigma}
\crcr
&+&
2\tau^z_{\sigma,\sigma}S^z\sum_{\vec k\vec k',\sigma}f^ \dag_{\vec k\eta\sigma}f_{\vec k'\eta'\sigma}
\Bigg],
\eeqa
where we have introduced the notation $\bar\sigma=-\sigma$.
Notice that only the first term contributes to the spin-flip processes. In particular, the term responsible of rising the spin of the local spin is given by
\beq
\sum_{\eta\eta'}  \frac{J_{\eta\eta'}}{2}
S^+ 
\sum_{\vec k\vec k',\sigma}f^ \dag_{\vec k\eta\downarrow}f_{\vec k'\eta'\uparrow}
\label{spin-flipK}
\eeq
Now we try to evaluate the equivalent term in the cotunneling Hamiltonian (\ref{Hcotun_eff}). The cotunneling Hamiltonian can be written as\footnote{Since we are interested only in the spin-flip processes, we can omit the  correction term in $H_{\rm cot.}'$ diagonal in the electrodes degrees of freedom.}
\beq
H_{cot.}\approx \sum_{\alpha\alpha'}
\left[\hat{T}_+(\alpha\alpha')-\hat{T}_-(\alpha'\alpha)\right]f_\alpha^\dag f_{\alpha'}
\label{Hcotun_effP}
\eeq
Now we make the following approximation. Since the scattering is dominated by processes close to the Fermi surface, we approximate the virtual transition operators $\hat T^\pm$ by their values at the Fermi level. This approximation is consistent with the expression of the Kondo exchange Hamiltonian $V_K$, where we have neglected the variation of the exchange coupling with the wave vectors. Then, we split the cotunneling Hamiltonian into two contributions, a spin-conserving and the spin-flip term. The last one is given by (after using the explicit quantum numbers)
\beq
H_{\rm cot.}^ {\rm sf}=\sum_{\eta\eta'}\sum_{\sigma}\left[\hat{T}_+(\eta\sigma,\eta'\bar\sigma) -
\hat{T}_-(\eta'\bar \sigma,\eta\sigma)
 \right]
 \sum_{\vec k\vec k'}f^{\dagger}_{\vec k\eta\sigma}f_{\vec k'\eta'\bar\sigma}, 
% -\sum_{\alpha}\left
\label{spin-flipC}
\eeq
where we have omitted the wave-vector index to indicate the values evaluated at the Fermi surface.
A direct comparison of (\ref{spin-flipK}) and the equivalent term in  (\ref{spin-flipC}) allows us to identify the Kondo-exchange coupling. For that, we project over the states $|n\rangle$ and $|n'\rangle$ of the local spin (QS) connected by the Kondo coupling (cotunneling Hamiltonian)
\beq
J_{\eta\eta'}=\frac{2}{S^ +_{nn'}}\left[
\hat{T}_+^{nn'}(\eta\downarrow,\eta'\uparrow) -
\hat{T}_-^{nn'}(\eta'\uparrow,\eta\downarrow)
\right],
\label{Jexccot}
\eeq
where $\hat T^{nn'}_\pm\equiv \langle n|\hat T_\pm |n'\rangle$.
This expression permits us to extract the effective exchange coupling from the cotunneling Hamiltonian. The explicit expressions of the virtual transition operators $\hat T^{nn'}_\pm(\alpha,\alpha')$ can be found in Appendix
~\ref{appendixTpm}.

Notice that either a variation of the charging energies $E_{m_\pm}-E_0$ or a variation of the hybridization constants $V_{\eta,\ell}$ will automatically reflect into a variation of the exchange constants $J_{\eta\eta'}$.

Now we come back to the general Kondo Hamiltonian where, in addition to the exchange coupling proportional to $\vec  S \cdot \vec\sigma (\vec r_0)$, we may have a direct scattering term
\begin{eqnarray}
V_K=V_K^{\rm ex}+\sum_{\eta\ne \eta'}T_{\eta\eta'} I_S \otimes \hat N(\vec r_{0}),
\end{eqnarray}
where we have defined $T_{\eta\eta'}=J_{\eta,\eta'}\sqrt{\xi}$. Here $I_S$ is the identity in the Hilbert space  of the local spin, and $\hat N(\vec r_{0})=\sum_{\alpha} f_\alpha^\dag f_\alpha$. 
Following an analysis similar to the one used for the spin-flip terms, and taking special care to split the spin-independent scattering and the spin-elastic scattering, we can arrive to the expression (see Appendix
 \ref{DSKondocot} for details):
\beq
T_{TS}=\left( \hat T^+_{nn}(E_F,TS,{\uparrow})-\hat{T}^-_{nn}(E_F,ST,\uparrow) \right)-
 \frac{J_{TS}}{2}  S_{nn}^z .
\label{relationTJf}
\eeq

It is worth pointing out that the cotunneling description and the Kondo exchange model with the parameters extracted from Eqs. (\ref{Jexccot}) and (\ref{relationTJf}) are not fully equivalent. This can be easily checked by looking at the result for the case of a transition metal adatom with the $d^1$ configuration, the $S=1/2$ counterpart, presented in Appendix
~\ref{TiH-case}. The first important difference with the single orbital Anderson model, which can be mapped to the $S=1/2$ Kondo exchange model, is that the coupling with the reservoirs can occur through different orbitals. While in the case of the coupling to the STM tip it may be a reasonable ansatz, this may not be the case for the substrate coupling. If the coupling to the reservoir occurs only through one orbital state, the $\hat T_+(\sigma\sigma')$ can be written as
\beqa
-i\hat{T}_-(\downarrow,\uparrow)&=&\gamma_{TS}^- |\uparrow\rangle\langle \downarrow|,
\crcr
\hat{T}_-(\downarrow,\downarrow)&=&\gamma_{TS}^-|\uparrow\rangle\langle \uparrow|,
\eeqa
where $\gamma_{TS}^-=-V_T V_{S}/\mu_- \ge 0$. This 
result is  the same that in the single Anderson impurity model obtained through the Schrieffer and Wolff transformation. 
Then, a second and more subtle difference appears in the virtual transition operator $\hat T_+(\sigma\sigma')$. Contrary to the single orbital Anderson model,  {\em the lowest energy states of the $d^2$ configuration correspond to a spin triplet} ($S=1$), with $L\approx 2$ where the two electrons are located on different orbitals with parallel spins. Crucially, these three triplet states $|t_1\rangle,|t_0\rangle$, and $|t_+\rangle$ do not contribute to the spin-flip terms, nor to the potential scattering between tip and surface.
Thus, we will need the next excited state for the transport properties, which is a spin and orbital singlet state, $|s\rangle=|0,0\rangle$.  
A direct inspection of the transition matrix elements of $\hat T_+(\alpha,\alpha)$ leads to the following result:
\beqa
i\hat{T}_+(\downarrow,\uparrow) &\approx& i\gamma_{TS}^+ d^\dag_{0,\downarrow}|s\rangle\langle s|d_{0,\uparrow}
\approx \gamma_{TS}^+ |\downarrow\rangle\langle\uparrow|
\label{TplusS_InMT}
\crcr
i\hat{T}_+(\downarrow,\downarrow)&\approx& i\gamma_{TS}^+ d^\dag_{0,\downarrow}|s\rangle\langle s|d_{0,\uparrow}
\approx \gamma_{TS}^+ |\downarrow\rangle\langle \downarrow|.
\eeqa
These results are qualitatively analogous to those obtained for the single orbital Anderson model, see Eqs. (\ref{TmAndsf}) and (\ref{TmAndsc}) in Appendix
~\ref{TiH-case}.  
Here we have neglected the denominator's energy differences, considering that the typical addition and removal energies are of the order of 1 eV or larger. Moreover, we have neglected contributions from other virtual states, which can be of the order of a 10\%. 

 Finally, these results deserve additional comments. First, the approximations involved in (\ref{TplusS_InMT}) do not hold in general. They are based on the assumption that tip and substrate couplings are only with the $d_{z^2}$ orbital of the transition metal atom. In addition, the specific properties $|t_1\rangle,|t_0\rangle$, and $|t_+\rangle$, which leads to a negligible contribution of the (virtual) lowest energy triplet, are specific of the symmetry considered. Second, the cotunneling description includes effects that are not considered within the standard exchange coupling description, mainly the charge fluctuations.
  For instance, the spin Hamiltonian dressing due to a Kondo exchange coupling can only affect the spin spectrum when the reservoir is spin-polarized, playing the role of an effective magnetic field, as in Eq. (\ref{Befft}). 
  However, the cotunneling Hamiltonian includes a finite correction $\Delta {\cal H}_S$ even for the coupling to a spin-averaging electrode. This has important consequences for the STM-ESR since, as described in Ref.~(\cite{Galvez_Wolf_prb_2019}), it can explain the DC current-independent Rabi frequency observed for some STM tips~\cite{Willke_Paul_sciadv_2018}.
 
Recent calculations based on the cotunneling physics, computing the full time-dependent current through an atomic orbital in the presence of atomic magnetic transitions \cite{Jose}
shows that the modulation of the hopping amplitude in time, or equivalently, the modulation of the tunnel barrier by the driving electric field, is
far enough to reproduce the STM-ESR spectra as a function of driving frequency. As in the above theory, the problem %to be determined 
is the degree
of actual change of the barrier by the driving field. 
Although in the simple description used in Eq. (\ref{tausb}), we have assumed that the RF modulation only affects
 the barrier height, a more precise picture would also require to account for the barrier width 
variation associated with the piezoelectric displacement discussed in Sec.~\ref{piezoR}. Nevertheless, both effects will give place to a modulation of the hopping amplitudes $V_{\alpha,{\bf i}}$, Eq. (\ref{Vexpbarr}),  as considered in the cotunneling picture.
Our calculations seem to indicate that sizeable barrier modulations are achievable with usual STM-induced fields, leading us to the upbeat conclusion that STM-ESR should transcend the use of MgO as the sole STM-ESR substrate.

\subsection{Other possible origins}
 In addition to the physical mechanisms signaled above, several alternative explanations deserve some comments. Although under the current experimental conditions of STM-ESR, the present experimental data suggest they may have a negligible contribution, they could be relevant under different experimental conditions. A novel mechanism was proposed by Shakirov {\em et al.}~\cite{Shakirov_Rubtsov_prb_2019}, who suggested that 
the resonance effect is a consequence of the spin-transfer-torque (STT), i.e., the nonlinearity of the
coupling between the magnetic moment and the spin-polarized current. 
This nonlinear process is enhanced near resonance, and it is proportional to the square of the driving amplitude $V_{RF}$. Moreover, 
the effect does not rely on any mechanical or orbital degrees of freedom, involving only the dissipative interaction with the spin-polarized tunnel current, which agrees well with the observed ubiquity of the resonant effect. 
 
When compared to the experimental findings, the peak resonance signal was found to be proportional to the DC current and to saturate with the RF bias voltage~\cite{Willke_Paul_sciadv_2018}, in contrast to the quadratic dependence expected from the STT mechanism.  Moreover, the strong dependence of the Rabi frequency with the standpoint distance $d_0$ at constant DC setpoint current does not seem compatible with this mechanism~\cite{Seifert_Kovarik_eabc_2020}. 

The modulation of the tunnel barrier by the exchange interaction between the localized spin and the tunneling electrons was already suggested~\cite{Manassen_Balatsky_ijc_2004,Balatsky_Nishijima_advphys_2012} as a possible origin of the resonance signal after the first STM-ESR attempts. Although this mechanism can be related to the exchange mechanism of Sec.\ref{JoaquinT} and the cotunneling mechanism of Sec.~\ref{cotSec}, the theoretical treatment was based on the analysis of the elastic current. It was argued that the barrier height for a particular electron scattering has a contribution coming from magnetic coupling, which depends on the average spin. This gives place to a modulation of the current that depends on the relative orientation between the localized spin and
the tunneling electron's spin. Then, they adduced that tunneling electrons keep the memory of
the spin polarization on a finite time scale, leading to a nonzero instantaneous spin current that could be related to the 1/$f$ noise features~\cite{Manassen_Balatsky_ijc_2004}. 

The modulation of the tunnel barrier is also behind a class of proposed physical mechanisms where the relativistic effects are essential for magnetic resonance. Balatsky {\em et al.} proposed a mechanism where the modulation of the tunnel barrier was mediated by the spin-orbit interaction of the substrate-metal
conduction electrons~\cite{Balatsky_Martin_qip_2002}. This effect translates into a current-independent modulation of the density of states by the precessing spin, which should work even in the absence of current spin-polarization,  in contrast with the broadly accepted STM-ESR measurements~\expSTMESR. 
The modulation of the tunneling barrier by the electric dipole moment originating from the spin-orbit interaction was also considered by Shachal and Manassen~\cite{Sachal_Manassen_prb_1992}, but this effect was later estimated to be negligible~\cite{Prioli_Helman_prb_1995}.
Other mechanisms mediated by relativistic effects were the coherent interference of the two resonance tunneling-current components passing
through the Zeeman levels split by the static magnetic field~\cite{Mozyrsky_Fedichkin_prb_2002},
spin-flip coupling between the conduction electrons on the localized spin site and electrodes~\cite{Zhu_Balatsky_prl_2002},  
%Creo que este coincide con el nuestro en esencia
%
coherent tunneling of a pair of electrons with opposite spins~\cite{Molotkov_sc_1994} among others~\cite{Balatsky_Nishijima_advphys_2012}.

\section{{Concluding remarks}}

The growing number of experimental results on STM-ESR is starting to delineate the role of the different experimental variables that control the resonance signal. They have clarified the relevance of the DC and RF bias voltage, the tunnel conductance, the temperature, or some of the tip microscopic details~\cite{Willke_Paul_sciadv_2018,Seifert_Kovarik_eabc_2020,Weerdenburg_Steinbrecher_arxiv_2020}. 

A clear picture emerges where the coupling with the spin-polarized tip plays a key role. STM-ESR can be observed even in the absence of external static magnetic field~\cite{Willke_Singha_nanolett_2019,Steinbrecher_Weerdenburg_arxiv_2020}, but the tip polarization is indispensable. \fer{The resonance field-dependence with the tip-adatom standoff distance also evinces direct interaction with the effective field produced by the tip magnetic moment~\cite{Seifert_Kovarik_eabc_2020}. }
 Once the magnetic field associated with the RF electric field has been discarded due to its tiny magnitude~\cite{Yang_Paul_prl_2019,Seifert_Kovarik_pr_2020}, the tip-induced magnetic field appears as the most likely source of coupling between the RF electric field and the local spin.  
The data point at two sources of tip-induced magnetic field~\cite{Seifert_Kovarik_eabc_2020}, a dipolar one
due to the direct interaction between the dipolar moment of the adatom and of the magnetic tip, and an exchange-field one that is shorter range. The exchange field has been argued to be sufficient to explain STM-ESR on
Fe adatoms~\cite{Lado_Ferron_prb_2017}, but recent data suggest that the longer-range contribution 
of the dipolar field is needed~\cite{Seifert_Kovarik_eabc_2020}. The exchange field source is ultimately
the Kondo exchange that is due to the overlap of the tip and adatom electronic wave function. As such, the
tunneling barrier between tip and adatom becomes a fundamental ingredient to induce sizeable exchange fields.

There is s growing consensus on the need to have a tip-induced
magnetic field, as we just saw, but one mandatory ingredient is needed; a time-dependent field. Indeed, ESR requires an RF oscillating magnetic field acting on the magnetic adatom. Several related mechanisms could explain this oscillating field. On one side, different works point to the piezoelectric displacement of the adatom with the RF electric field. Estimations based on density functional calculations combined with multiplet calculations shows in general qualitative agreement with the experimentally observed Rabi frequency~\cite{Lado_Ferron_prb_2017, Yang_Paul_prl_2019,Seifert_Kovarik_eabc_2020}. The differences, which can vary from a factor 2-4 \cite{Seifert_Kovarik_eabc_2020,Lado_Ferron_prb_2017} to almost a factor 40~\cite{Yang_Paul_prl_2019}, have been attributed to the harmonic approximation in the estimation of the recovery constant~\cite{Seifert_Kovarik_eabc_2020} or to the piezoelectric motion of the MgO layer, as well as the
motion of the Fe atom at the tip apex~\cite{Yang_Paul_prl_2019}. Independent of the other piezoelectric displacement sources, it is clear that this mechanism can drive an RF magnetic coupling with the local spin. 
However, the cotunneling proposal~\cite{Galvez_Wolf_prb_2019} shows that a piezoelectric response is not needed
\fer{as} long as the RF electric field strongly modifies the tunneling probability for the electron. In this
case, the tunneling barrier modulation leads to an effective time-dependent magnetic field that
suffices to induce Rabi oscillations. The piezoelectric effect enhances the barrier modulation, facilitating the
appearance of the magnetic
resonance. For these reasons, we believe that STM-ESR should be feasible on different types of substrates
other than MgO/Ag(100).

\section*{Acknowledgements}
We are pleased to thank our collaborators for important discussions. A non-exhaustive list of the
many contributors to our discussions is: 
D.-J. Choi, 
T. Choi, 
F. Donatti,
A. J. Heinrich,
J. Fern\'andez-Rossier,
P. Gambardella,
L. Limot,
F. Natterer,
J. Reina G\'alvez,
T. Seifert,
S. Stepanow,
P. Willke,
C. Wolf.
Financial support from the Spanish MICINN (projects RTI2018-097895-B-C44, 
PlD2019-109539G8-C41, \fer{PID2019-109539GB-C41}) is gratefully acknowledged.
FD acknowledges financial support from Basque Government, grant IT986-16
and Canary Islands program {\em Viera y Clavijo}
(Ref. 2017/0000231).

\appendix
\renewcommand{\thesection}{\Alph{section}}

%%%%%%%%%%%%%         Appendix Ti-H: cotunneling vs. Kondo exchange		%%%%%%%%%%%%%%   
\section{Virtual transition operators for the cotunneling model \label{appendixTpm}}

In the case of a multiorbital correlated QS weakly coupled to the electrodes, the matrix elements of the virtual transition operators $\langle n|\hat{T}_\pm(\alpha\alpha')|n'\rangle$ between the eigenstates of ${\cal H}_S$ are given by~\cite{Delgado_Rossier_prb_2011,Galvez_Wolf_prb_2019}:
\beqa
\langle n|\hat{T}_\mp(\alpha\alpha')|n'\rangle &=&  
\sum_{m_-,\ell\ell'}\frac{V^*_{\alpha,\ell}V_{\alpha',\ell'} }{E_{m_-}-E_0\pm \epsilon_{\alpha} }
\gamma_{nn'}^{m_\mp}(\alpha \ell,\alpha' \ell')
\crcr
&& \label{Tpm}
\eeqa
where $m_\mp$ labels the eigenstates of ${\cal H}_S$ with $N\mp 1$ electrons, $E_0$ is the ground state energy of ${\cal H}_S$, $\epsilon_\alpha$ denotes the quasiparticle energy, and 
 $\gamma_{nn'}^{m^-}(\alpha\ell',\alpha\ell)=\langle
n|d_{\ell\sigma}^\dag |m_-\rangle \langle m_-|d_{\ell'\sigma}|n'\rangle$
and $\gamma_{nn'}^{m^+}(\alpha\ell',\alpha\ell)=\langle n|d_{\ell\sigma}
|m_+\rangle \langle m_+|d_{\ell'\sigma}^\dag|n'\rangle$.\footnote{For clarity reasons, here we have assumed real coupling matrix elements $V_{\alpha,{\bf i}}$} Here $d_{\ell,\sigma}$ and $d_{\ell,\sigma}^\dag$ denotes the annihilation and creation operators in the $\ell$ orbital of the QS with spin $\sigma$.

\section{Derivation of the direct scattering term \label{DSKondocot}}
In the case of the elastic scattering, we need to include not only the off-diagonal terms of the cotunneling Hamiltonian but also the diagonal ones. This makes it necessary to include the corrected cotunneling Hamiltonian $H_{\rm cot.}'$ instead of $H_{\rm cot.}$ since the diagonal terms will also contribute to the spin-conserving processes.
First, we project on the states $|n\rangle\otimes |r_{\eta,\sigma}\rangle$, where $|r_{\eta,\sigma}\rangle$ denotes an eigenstate of the reservoir $\eta$ with spin $\sigma$ and energy $\epsilon_r$ 

For the cotunneling Hamiltonian $H_{\rm cot.}'$, the matrix elements will be given by
\beqa
\langle n|\otimes \langle r_{\eta,\uparrow}|H_{\rm cot}'| n'\rangle\otimes |r'_{\eta',\uparrow}\rangle 
&=&   \langle r_{\eta,\uparrow}|f_\alpha^\dag f_{\alpha'}|r'_{\eta',\uparrow}\rangle
\crcr
&&\hspace{-5.cm}\times
 \sum_{\alpha\alpha'}\langle n\left[\hat T_+(\alpha\alpha')-\hat{T}_-(\alpha'\alpha)
\right]|n'\rangle - \langle r_{\eta,\uparrow}|\hat n_{\alpha} |r'_{\eta',\uparrow}\rangle
\crcr
&&\hspace{-5.cm}\times\sum_{\alpha}\langle n|\left[\hat T_+(\alpha\alpha)-\hat{T}_-(\alpha\alpha)
\right]|n'\rangle. 
\eeqa
As we did for the spin-flip terms, we consider that the scattering processes are mainly due to the vicinity of the Fermi surface and, besides, we also assume that the hoppings $V_{\alpha,{\bf i}}$ do not depend on the direction of the wave-vector $\vec k$. Moreover, we particularize the previous expressions for the case $n=n'$ and $\sigma=\uparrow$, so that
\beqa
\langle n|\otimes \langle r_{\eta,\uparrow}|H_{\rm cotunn}'| n\rangle\otimes |r'_{\eta',\uparrow}\rangle 
&=&\sum_{\vec k\vec k'}\langle r_{\eta,\uparrow}|f_{\eta_{\vec k\uparrow}}^\dag f_{\eta'_{\vec k'\uparrow}}|r'_{\eta,\uparrow}\rangle
\crcr
&&\hspace{-5.cm}\times \left[\hat T^+_{nn}(\eta_{\uparrow},\eta'_{\uparrow})-\hat{T}^-_{nn}(\eta'_{\uparrow},\eta_{\uparrow})
\right]-\delta_{\eta\eta'}\sum_{\vec k}\langle r_{\eta,\uparrow}|f_{\eta_{\vec k\uparrow}}^\dag f_{\eta_{\vec k\uparrow}} |r'_{\eta,\uparrow}\rangle
\crcr
&&\hspace{-5.cm}\times\left[\hat T^+_{nn}(\eta_{\uparrow},\eta_{\uparrow})-\hat{T}^-_{nn}(\eta_{\uparrow},\eta_{\uparrow})
\right] ,
 \label{ProjectCotP}
\eeqa

We now evaluate the spin-conserving terms of $V_K$. As for the cotunneling Hamiltonian, we look at the matrix elements of the Kondo Hamiltonian and the states 
$|n\rangle\otimes |r_{\eta,\sigma}\rangle$. Then, we get
\beqa
\langle n|\otimes \langle r_{\eta,\uparrow}|V_K'|n'\rangle\otimes|r'_{\eta',\uparrow}\rangle 
&=&\left(
 \frac{J_{\eta\eta'}}{2}  S_{nn'}^z +\delta_{nn'}T_{\eta\eta'} \right) 
\crcr
&&\hspace{-1.5cm}\times
\sum_{\vec k\vec k'}\langle r_{\eta\uparrow}| f_{\vec k\eta\uparrow}^\dag f_{\vec k'\eta'\uparrow}|r'_{\eta'\uparrow}\rangle,
\label{projectVK}
\eeqa
where we have introduced the short notation $\hat T^\pm_{nn'}(\eta\sigma,\eta'\sigma')\equiv \langle n|\hat T_\pm (k_F\eta\sigma,k_F\eta'\sigma')|n'\rangle$. A direct comparison between expressions (\ref{projectVK}) and ({\ref{ProjectCotP}) allows us to identify the following relation:
\beq
T_{TS}=\left( \hat T^+_{nn}(E_F,TS,{\uparrow})-\hat{T}^-_{nn}(E_F,ST,\uparrow) \right)-
 \frac{J_{TS}}{2}  S_{nn}^z ,
% -\delta_{\eta\eta'}\left( \hat T^+_{nn'}(\eta_{\uparrow},\eta_{\uparrow})-\hat{T}^-_{nn'}(\eta_{\uparrow},\eta_{\uparrow})\right)
\label{relationTJfApp}
\eeq
which is the equation used in the main text.

\section{Example of an $d^1$ electronic configuration: the TiH on MgO (bridge position) \label{TiH-case}}
We now calculate the transition matrix elements obtained from the multiplet calculation for the TiH $d^1$ adatom on the bridge side of MgO/Ag(100). In that system, the lowest energy orbital state corresponds to the $d_{z^2}$ orbital ($m_L=0$).  We thus assume that only this  $d_{z^2}$ orbital is hybridized with the surface and tip electrodes, with intensities $V_S$ and $V_T$, respectively (we assume that those couplings are real). In addition, we evaluate the different operators only at the {Fermi level}, i.e., $k=k'=k_F$.  

 We first start calculating the expressions of the $\hat T_-$ operator and we particularize to the tip-substrate tunneling term. For clarity, we omit the electrode indexes whenever possible. Considering that the vacuum state has only one state, $|0\rangle$, we have that
\begin{eqnarray}
\hat{T}_-(\sigma,\sigma')=
-\frac{V_T V_{S}}{\mu_-}
d^{\dagger}_{0,\sigma}|0\rangle
\langle 0 |d_{0,\sigma'},
\end{eqnarray} 
where we have defined $\mu_-=E_0-E_{0_-}-E_F$. 
\fer{Since there is no magnetic anisotropy for the neutral charge space, we have the two spin states $\sigma=\pm 1/2$.}
Notice that, since the orbital state is the (non-degenerate) ground state in the presence of the crystal field can be written as $|\sigma,0\rangle$, where the $0$ label denotes the $d_{z^2}$ orbital state. Hence, we have for the spin-flip terms the following operators in the $\{|\uparrow\rangle,|\downarrow\rangle\}$ bases:
\beqa
-i\hat{T}_-(\downarrow,\uparrow)=\gamma_{TS}^-\left(
\begin{array}{cc}
0 & 1\\
0 &	0
\end{array}
\right),
\label{TmAndsf}
\eeqa
and the spin conserving term
\beqa
\hat{T}_-(\downarrow,\downarrow)=\gamma_{TS}^-\left(
\begin{array}{cc}
1 & 0\\
0 &	0
\end{array}
\right).
\label{TmAndsc}
\eeqa
(Similarly for the $\uparrow\uparrow$ matrix elements). 

We now consider the  $\hat T_+$ operator. In the first place, we remind the results obtained for the single Anderson impurity model. In that case, the charged state with two electrons consists of a doubly occupied singlet state $|\uparrow\downarrow\rangle$. Then, one gets the following transition operators
\beqa
\hat{T}_+^{\rm Ander}(\downarrow,\uparrow)=\gamma_{TS}^+\left(
\begin{array}{cc}
0 & 0\\
1 &	0
\end{array}
\right),
\eeqa
and the spin conserving term
\beqa
\hat{T}_+^{\rm Ander}(\downarrow,\downarrow)=\gamma_{TS}^+\left(
\begin{array}{cc}
0 & 0\\
0 &	1
\end{array}
\right),
\eeqa
where $\gamma_{TS}^+=V_T V_{S}/\mu_+ \ge 0$, with $\mu_+=E_{0_+}-E_0-E_F$. Let us come back to the  TiH $d^1$ system. 
\fer{Thus, we will need the next excited state for the transport properties, which is a spin and orbital singlet state, $|s\rangle=|0,0\rangle$. }
Neglecting the energy differences in the denominator, and contributions from other higher energy states, which could account for corrections of the order of 10\%,
we get that
\beqa
i\hat{T}_+(\downarrow,\uparrow)\approx i\gamma_{TS}^+ d^\dag_{0,\downarrow}|s\rangle\langle s|d_{0,\uparrow}
\approx \gamma_{TS}^+
\left(
\begin{array}{cc}
0 & 0\\
1 &	0
\end{array}
\right).
\label{TplusS_In}
\eeqa
 For the spin-conserving terms, we find that
\beqa
i\hat{T}_+(\downarrow,\downarrow)\approx i\gamma_{TS}^+ d^\dag_{0,\downarrow}|s\rangle\langle s|d_{0,\uparrow}
\approx \gamma_{TS}^+
\left(
\begin{array}{cc}
0 & 0\\
0 &	1
\end{array}
\right).
\label{TplusS_El}
\eeqa
With this expressions we can arrive to the following relations:
\beq
J_{T,S}\approx \frac{2}{S^ +_{nn'}}\left(
 \gamma_{TS}^- +\gamma_{TS}^+ 
 \right),
\eeq
and
\beq
T_{TS} \approx \gamma_{TS}^+ 
 \frac{J_{TS}}{2}  S_{nn}^z 
 \approx \gamma_{TS}^+ +\frac{S^z_{nn}}{S^ +_{nn'}}\left(
 \gamma_{TS}^- +\gamma_{TS}^+ 
 \right).
\eeq
If we particularize  to the case of electron-hole symmetry, where $\gamma^{-}_{TS}=\gamma_{TS}^+\equiv \gamma_{TS}$, and considering that $S^z_{-,-}=-1/2$ and $S^ +_{+,-}=1$, as it corresponds to an isotropic $S=1/2$ spin system in a field $B_z>0$, we get that
\beq
J_{T,S}^{eh}\approx 4\gamma_{TS}\approx\frac{8V_{T}V_S}{U}
\eeq
and
\beq
T_{TS}^{eh} 
 \approx \gamma_{TS} \left(1 -\frac{1}{2}\times 2
 \right)= 0.
\eeq
These last two equations correspond to the Schrieffer and Wolff transformation results for the $S=1/2$ spin system~\cite{Schrieffer_Wolff_pr_1966}.

%%%%%%%%%%%%%          Bibliography		%%%%%%%%%%%%%%   
%\bibliographystyle{apsrev}
%\bibliographystyle{elsarticle-num} 
%\bibliography{ref_01_09_2020}

\end{document}